\documentclass[journal]{IEEEtran}
\IEEEoverridecommandlockouts

\usepackage[utf8]{inputenc}
\usepackage{cite}
\usepackage{floatrow}
\usepackage{amsmath,amssymb,amsfonts,stfloats}
\usepackage{tabularx}
\usepackage[left=0.64in,right=.64in,top=0.75in,bottom=1.1in]{geometry}
\usepackage{algorithmic}
\usepackage{graphicx}
\usepackage{textcomp}
\usepackage{floatrow}
\usepackage{xcolor}
\usepackage{booktabs}

\usepackage{multicol}
\usepackage{acronym}
\usepackage{url}
\usepackage{mathtools}
\usepackage[font=small]{caption}
\usepackage[caption=false,font=footnotesize]{subfig}
\usepackage{comment}
\def\BibTeX{{\rm B\kern-.05em{\sc i\kern-.025em b}\kern-.08em
    T\kern-.1667em\lower.7ex\hbox{E}\kern-.125emX}}

\newcommand{\linebreakand}{%
  \end{@IEEEauthorhalign}
  \hfill\mbox{}\par
  \mbox{}\hfill\begin{@IEEEauthorhalign}
}

\ifCLASSINFOpdf
\else
\fi
\begin{document}

\title{Sensing in NLOS: a Stroboscopic Approach
\thanks{This work was partially supported by the European Union under the Italian National Recovery and Resilience Plan (NRRP) of NextGenerationEU, partnership on “Telecommunications of the Future” (PE00000001 - program “RESTART”).}
\thanks{The authors are with the Department of Electronics, Information and Bioengineering, Politecnico di Milano, Via Ponzio 24/5, 20133 Milano, Italy.}
}

\author{ Davide~Tornielli~Bellini,~\IEEEmembership{Graduate~Student~Member,~IEEE}, Dario~Tagliaferri,~\IEEEmembership{Member,~IEEE}, Marouan~Mizmizi,~\IEEEmembership{Member,~IEEE}, Stefano~Tebaldini,~\IEEEmembership{Senior~Member,~IEEE}, Umberto~Spagnolini,~\IEEEmembership{Senior~Member,~IEEE}
}

\maketitle

\begin{abstract}
Sensing in non-line-of-sight (NLOS) is a well-known issue that limits the effective range of radar-like sensors. Existing approaches for NLOS sensing consider the usage of either metallic mirrors, that only work under specular reflection, or dynamically-reconfigurable metasurfaces that steer the signal to cover a desired area in NLOS, with the drawback of cost and control signaling. This paper proposes a novel sensing system, that allows a source to image a desired region of interest (ROI) in NLOS, using the combination of a proper beam sweeping (by the source) as well as a passive reflection plane configured as a periodic angular deflecting function (that allows illuminating the ROI). \textit{Stroboscopic sensing} is obtained by sweeping over a sufficiently large portion of the reflection plane, the source covers the ROI \textit{and} enhance the spatial resolution of the image, thanks to multiple diverse observation angles of ROI. Remarkably, the proposed system achieves a near-field imaging with a sequence of far-field acquisitions, thus limiting the implementation complexity. We detail the system design criteria and trade-offs, demonstrating the remarkable benefits of such a stroboscopic sensing system, where a possibly moving source can observe a ROI through multiple points of view as if it were static. 
\end{abstract}


\section{Introduction}

Radio sensing has been identified as one of the pivotal technologies to provide the environmental awareness expected for future wireless communication networks in the $3-300$ GHz band, such that to enable new application scenarios and market verticals (augmented and virtual reality, autonomous systems, and others)~\cite{9349624,9815783}. Typical device-free (radar-like) sensing allows detecting targets and tracking their motion, while gathering meaningful features, such as position, orientation, velocity and shape. In particular, \textit{localization} is the procedure to estimate the position, velocity and possibly orientation of targets~\cite{10287134,4802193}, while \textit{imaging} refers to the generation of a map of the reflectivity of the environment, from which to infer the number of targets (via detection) and their shape. Imaging performance is assessed in terms of resolution, i.e., the capability of distinguishing two closely spaced targets~\cite{manzoni2023wavefield,tagliaferri2023cooperative}.
Nowadays, device-free radio sensing is implemented by stand-alone radar systems, operating on dedicated bands with ad-hoc hardware platforms, for a broad remote sensing application domain, ranging from infrastructure monitoring, Earth observation, weather forecast to recent autonomous driving~\cite{TEBALDINI2023113532,rs14153602}. However, an ubiquitous deployment of radar sensing systems is incompatible with the required penetration of 6G networks since equipping each communication transceiver with a radar would lead to an unsustainable duplication of hardware resources. Therefore, integrating advanced radio sensing capabilities into existing and future network infrastructure requires exploiting the very same resources (time, frequency, and hardware) to both transfer information \textit{and} sense the environment. Integrated sensing and communication (ISAC) systems constitute a key research topic in light of 6G systems, in which ISAC is expected to be a cornerstone technology. A survey on the relevant research literature on ISAC systems can be found in~\cite{Liu_survey}.

In tandem with the development of radio sensing technology, recent technological advances concerning electromagnetic (EM) metamaterials gave rise to metasurfaces, i.e., 2D structures enabling advanced EM wave manipulations, such as anomalous reflection/refraction, diffusion, focusing, beam splitting and others~\cite{doi:10.1126/science.1210713}. At microwaves, metasurfaces are typically manufactured as 2D arrays of unit cells (meta-atoms), often referred to as \textit{EM skins} (EMSs). The reflection/transmission coefficients of meta-atoms can be tuned to obtained the desired functionality following the generalized Snell's law of reflection/refraction \cite{di2020smart,7109827}. When the aforementioned tuning changes in time, we refer to reconfigurable passive EM skins, also called reconfigurable intelligent surfaces (RISs). A RIS typically controls the phase of the reflection coefficient of each meta-atom by integrating a suitable circuitry in the single element, resulting in a quasi-passive technology. Along with applications for communication systems, RISs potential for sensing systems has been explored in a number of recent relevant literature, that mostly refers to localization rather than imaging~\cite{9775078}. The authors of~\cite{9508872} analyze the Cramér Rao bound (CRB) of position and orientation estimation of a RIS-aided localization system, proposing a suitable CRB-minimizing phase design approach. The works~\cite{Zhang2022metalocalization,Zhang2022metaradar} consider indoor user localization employing RIS and received signal strength measurements. Paper \cite{Wang2022_location_awareness} compares the localization performance of RIS-aided and continuous metasurface-aided ISAC systems, deriving the CRB in both cases. Other relevant works consider placing RISs (or EMSs) on targets, to enhance their reflectivity and consequently the localization performance \cite{tagliaferri2023reconfigurable,Wymeersch2021}.
Recently, localization studies involving RISs have been focused on the radiative \textit{near-field} effect, namely the operating condition for which the manipulated wavefront is curved across the RIS. In addition to the straightforward increase of the aperture and spatial resolution~\cite{9838638}, new degrees of freedom brought by near-field operation allow exploiting a large EMS/RIS in place of a large active antenna array, reducing cost and energy consumption \cite{9709801}. Position error bounds for near-field RIS-aided localization systems are derived and discussed in \cite{9625826}.

In addition to the aforementioned works, others explored the usage of RISs for NLOS sensing as well. Sensing in NLOS is a well-known issue that has been explored in various ways, see \cite{9468353,9553059,9547412,8966246}. All existing works make use of an a-priori knowledge of the surrounding geometry or a properly placed metallic mirror, that constrain the sensing system to exploit the specular reflections only. Moreover, the sensing resolution is dictated by the sensing platform, and it is usually limited by cost. EM skins, instead, enable anomalous reflections going beyond the specular one, extending the sensing capabilities, while RISs allow steering the reflected signal over a wide area in NLOS. 
In~\cite{Buzzi_RISforradar_journal}, the authors propose to use a RIS to assist radar in NLOS conditions, addressing the maximization of the probability of correct detection of a given target, via signal-to-noise (SNR) maximization. The CRB results  suggest to place the RIS near either the radar or the target, to combat the severe path loss induced by a double or triple reflection (Tx-RIS-target-RIS-Rx). Radar surveillance in NLOS scenarios assisted by RIS is proposed in \cite{9511765}, whereby different operating regimes of the RIS are discussed. The paper \cite{9650561}, instead, addresses the near-field target localization in NLOS scenarios, evaluating the CRB on position estimation and proposing an optimization problem to design the RIS phases for CRB minimization. These latter works address the usage of RIS for localization, but the literature concerning imaging is exiguous and mostly focused on biomedical use cases \cite{Ghavami2022_metasurfaces_radar}. Other relevant works are \cite{9299878,torcolacci2023holographic}. The authors of \cite{9299878} address the imaging problem as an inverse scattering reconstruction in far-field, considering distributed antenna systems. Recently, the authors of \cite{torcolacci2023holographic} tackled the imaging with RISs in both LOS and NLOS scenarios, designing the illumination pattern and the RIS coefficients to minimize the reconstruction error of the reflectivity map in a desired area of interest. Noteworthy, the image resolution in RIS-aided systems is dictated by the size of the RIS itself~\cite{9829750}, that is however impaired by high manufacturing costs, almost totally due to the circuitry
for dynamic phase configuration. 

\textit{Contribution}: This paper focuses on radio sensing in NLOS. We propose a novel imaging system for exploring a desired region of interest (ROI) in NLOS from a sensing terminal (source), exploiting the double reflection off a properly designed reflection plane. The source sweeps a sensing beam towards the plane, over a purposely designed Tx codebook, to illuminate multiple contiguous portions of the plane itself. The latter implements a periodic anomalous reflection function, that is jointly designed with the source's beam and Tx codebook to guarantee both the coverage of the ROI \textit{and} enhanced spatial resolution w.r.t. the source only capabilities. We detail the system design and performance for a moving imaging system, where both source and ROI move along a linear trajectory parallel to the SP-EMS. In this latter case, the proposed imaging system resembles a \textit{stroboscopic} imaging system to observe the ROI from the source as if it is static. This work generalizes our previous one \cite{bellini2024multiview}, that considers a fixed Tx beam and does not address the resolution analysis.
The contributions of the papers are listed in the following: 
\begin{itemize}

    \item We propose a novel stroboscopic system for sensing in NLOS, where a source aims at generating an image of a ROI through reflection from an anomalous plane that implements a periodic reflection pattern. This goes well beyond existing works on imaging in NLOS, e.g., see \cite{9511765,torcolacci2023holographic}, since we consider a low-cost static (non-reconfigurable) reflection plane that is purposely designed to maximize the image resolution as the main key performance indicator for the sensing system.  

    \item We detail the system design for a modular reflection plane. Each module is implemented with a static passive EMS (SP-EMS) and it is phase-configured according to far-field operation. SP-EMSs are non reconfigurable EMSs whose wave manipulation capabilities are defined in the design process, and come with orders-of-magnitude lower manufacturing costs than RISs \cite{9580737,tagliaferri2023reconfigurable}. We first derive the angular sampling limit at the source side to avoid spatial ambiguities within the ROI. Then, we quantitatively discuss the design criteria for the SP-EMS-based reflection plane (selection of the module size and of the periodicity of the reflection function). Noteworthy, we show that it is possible to design the reflection plane such that to image any target in the ROI with a spatial resolution comparable to the usage of the metasurface as a lens (maximum attainable resolution), demonstrating a substantial improvement compared to the usage of simple metallic mirrors. Remarkably, paper demonstrates how to achieve a near-field imaging result with successive far-field acquisitions, simplifying the design \cite{10541333}. The theoretical resolution analysis is addressed by means of diffraction tomography theory, and it applies to any sensing system exploiting a metasurface.    

    \item To avoid excessive abstract settings, we analyze, with both closed form derivations and numerical results, the effect of three main non-ideal settings for the imaging of the ROI, namely an error in the knowledge of the system setup. 
    
\end{itemize}

\textit{Organization}: The paper is organized as follows: Section \ref{sect:multiview} illustrates the concept of stroboscopicity in radio sensing, the system model is in Section \ref{sec:system_model}, Section \ref{sec:image_formation} describes the image formation algorithm and the required angular sampling, Section \ref{sec:system_design} details the system design, Section \ref{sec:nonidealities} addresses the effect of typical sources of errors in the image formation, Section \ref{sec:results} reports and discusses the simulation results while Section \ref{sec:conclusions} concludes the paper. 

\textit{Notation}: Bold lower-case letters stand for column vectors. With $\mathbf{a}\sim\mathcal{CN}(\boldsymbol{\mu},\mathbf{C})$ we denote a circularly complex multi-variate Gaussian random variable with mean $\boldsymbol{\mu}$ and covariance matrix $\mathbf{C}$. $\mathbb{R}$ and $\mathbb{C}$ denote, respectively, the set of real and complex numbers. $\delta_n$ is the Kronecker delta function.

\begin{figure}
    \centering
    \subfloat[][]{\includegraphics[width=0.7\columnwidth]{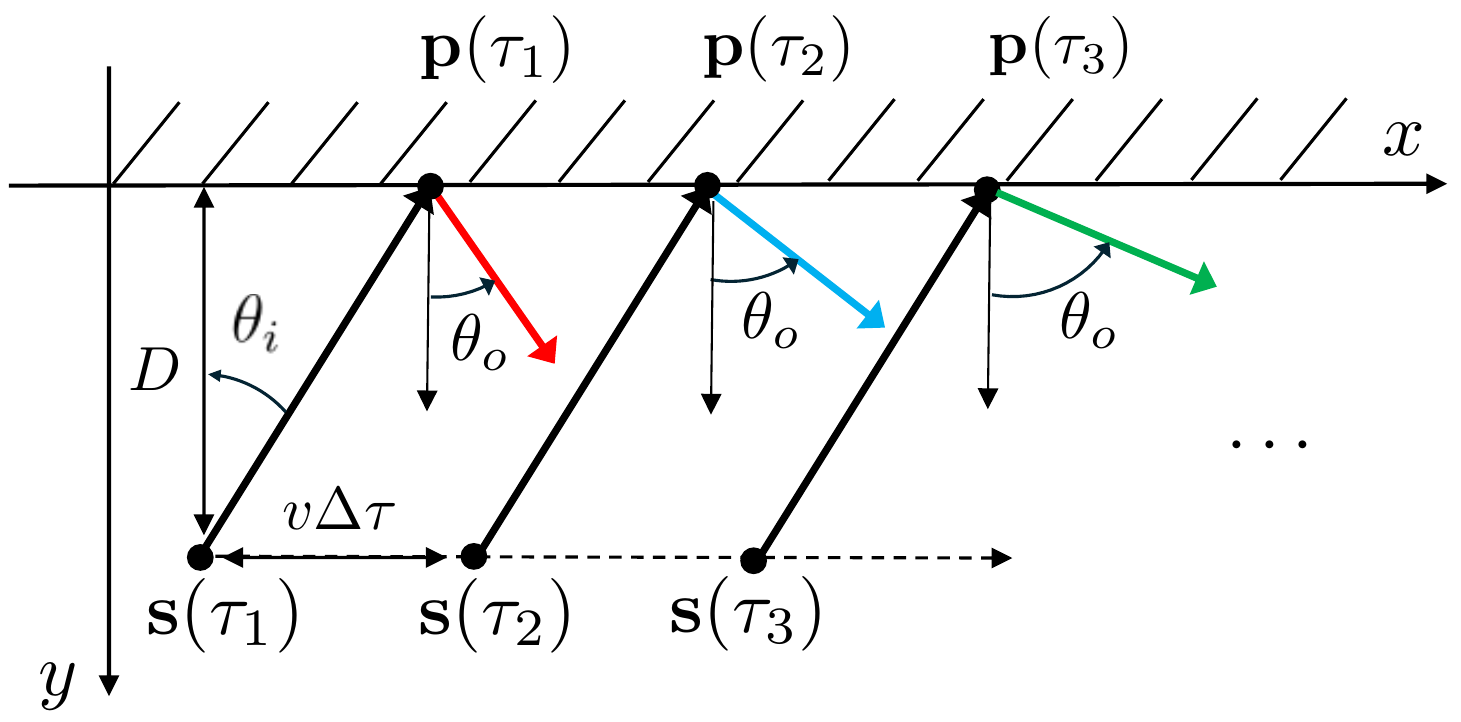}} \\ \vspace{-0.25cm}
    \subfloat[][]{\includegraphics[width=0.7\columnwidth]{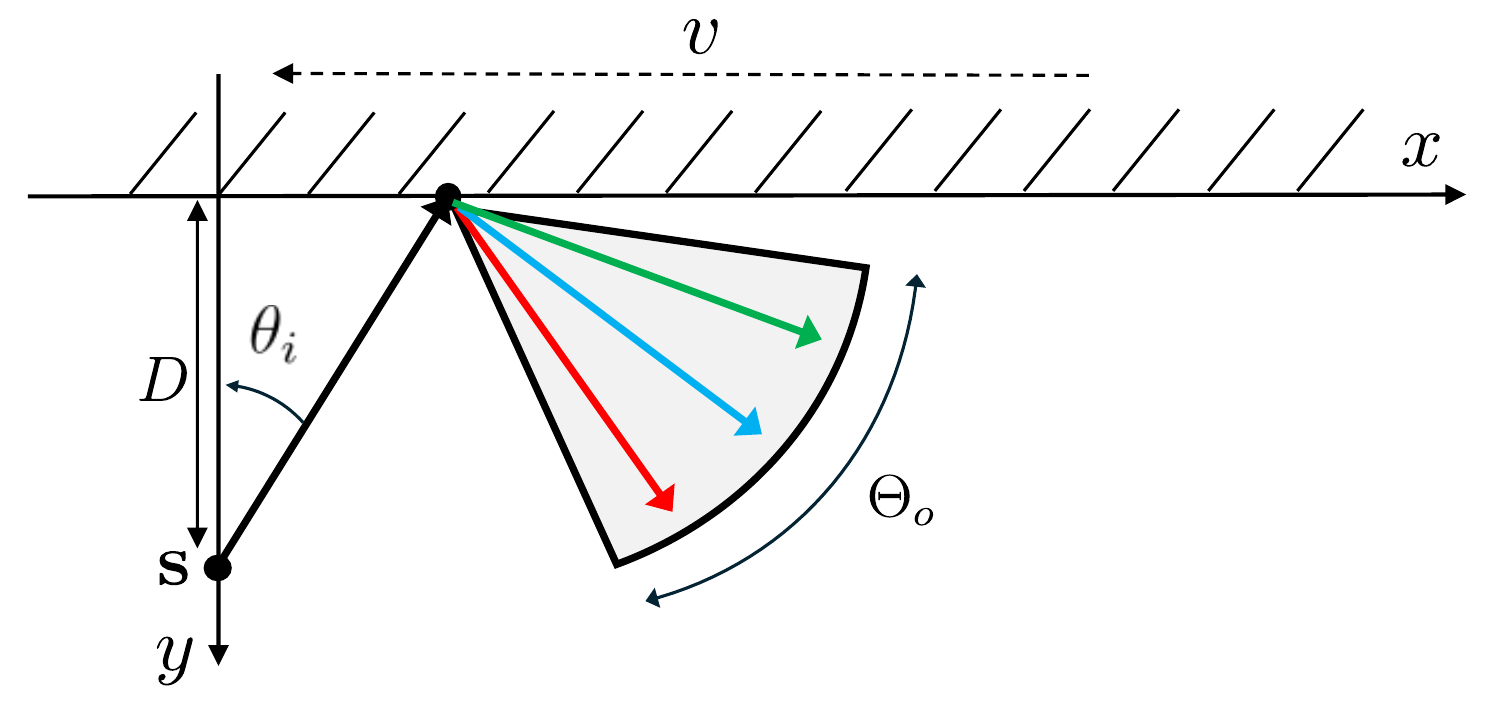}}
    \caption{Sketch of a system with a moving source and an anomalous reflection plane: (a) the source illuminates the plane with a fixed angle $\theta_i$, and the plane reflects according to the specific incident point $\mathbf{p}$; (b) the system (a) is equivalent to a fixed source and a reflection sweep over the angular span $\Theta_o$ activated by the moving reflecting plane. }
    \label{fig:fixed_inc_varying_refl}
\end{figure}

\section{Stroboscopic in Radio Sensing}
\label{sect:multiview}


Let us consider the 2D scenario depicted in Fig.~\ref{fig:fixed_inc_varying_refl}.
A source at distance $y = D$ is moving with speed $v$ along $x$, with a linear trajectory $\mathbf{s}(\tau) = (s_x(0) + v \tau, D)$. The source emits a repeated signal in direction $\theta_i$ once every $\Delta \tau$ seconds\footnote{The Tx signal can be a radar pulse or another specifically conceived waveform, e.g., an ISAC signal. We assume that the employed beamwidth at the source is narrow enough to approximate the emitted signal as a single ray in direction $\theta_i$.}, so that source position results sampled along the so called \textit{slow time} $\tau = \ell \Delta \tau$. Signal impinges on a reflection plane deployed along the $x$ axis at $y=0$ at point $\mathbf{p}(\tau) = (s_x(0) + D\tan \theta_i + \ell v \Delta \tau, 0)$. The reflection plane is capable of implementing an arbitrary reflection angle $\theta_o(x)$ that depends on the $x$ coordinate. This reflection angle follows a repeated pattern in space, periodically spanning an angular set $\Theta_o$.
Since the source emits a periodic pulse in a constant angle $\theta_i$ and moves at constant speed $v$ along $x$, the dynamic system is equivalent to a system where the source is fixed, $\mathbf{s}=(0,D)$, and the reflection plane moves at speed $v$ along $-x$. In the latter setup, the source is able to explore all the region within reflection angles $\theta_o(\tau) \in \Theta_o$ by emitting a sufficient number of pulses. A full sweep coverage can be achieved with a single incidence angle $\theta_i$, as illustrated in Fig.~\ref{fig:fixed_inc_varying_refl}b.


\begin{figure}[t]
    \centering
    \subfloat[][]{\includegraphics[width=0.7\columnwidth]{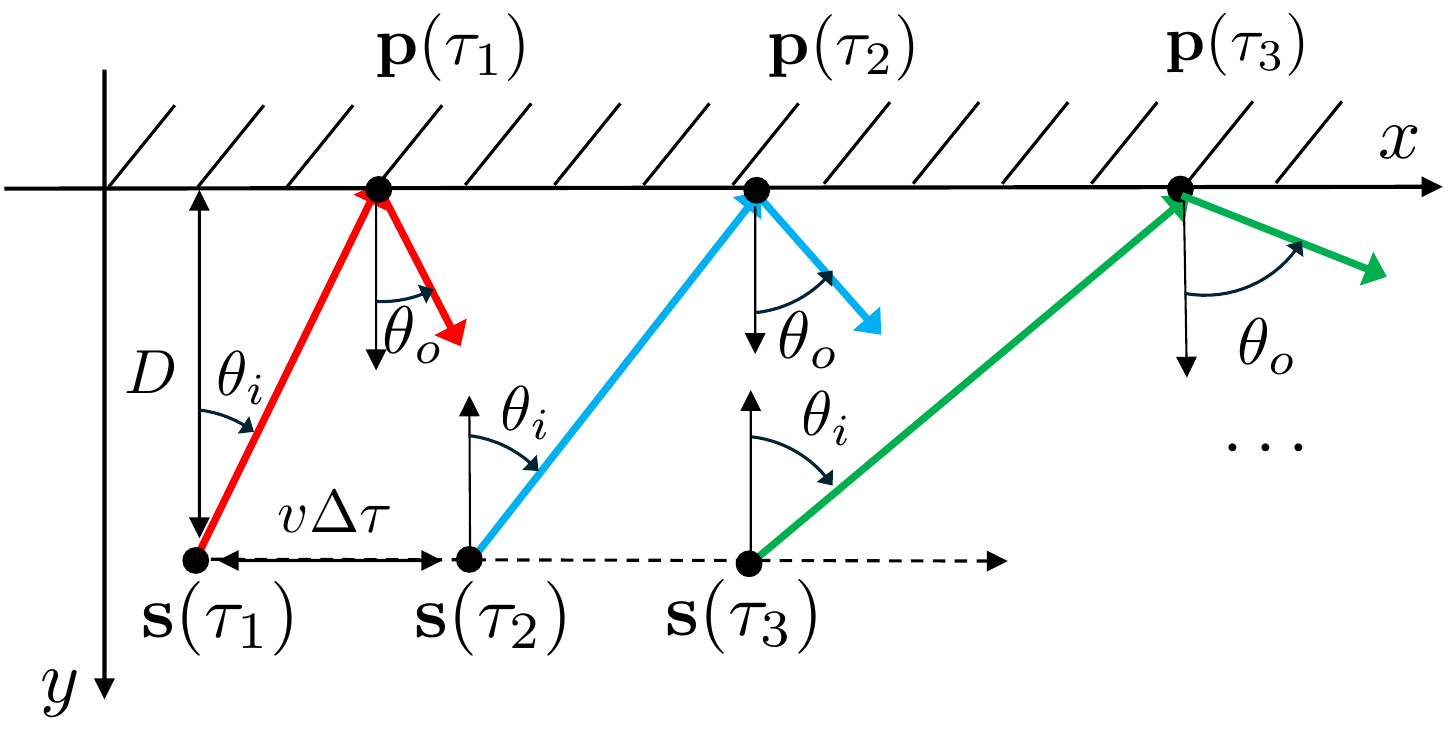}} \\ \vspace{-0.25cm}
    \subfloat[][]{\includegraphics[width=0.7\columnwidth]{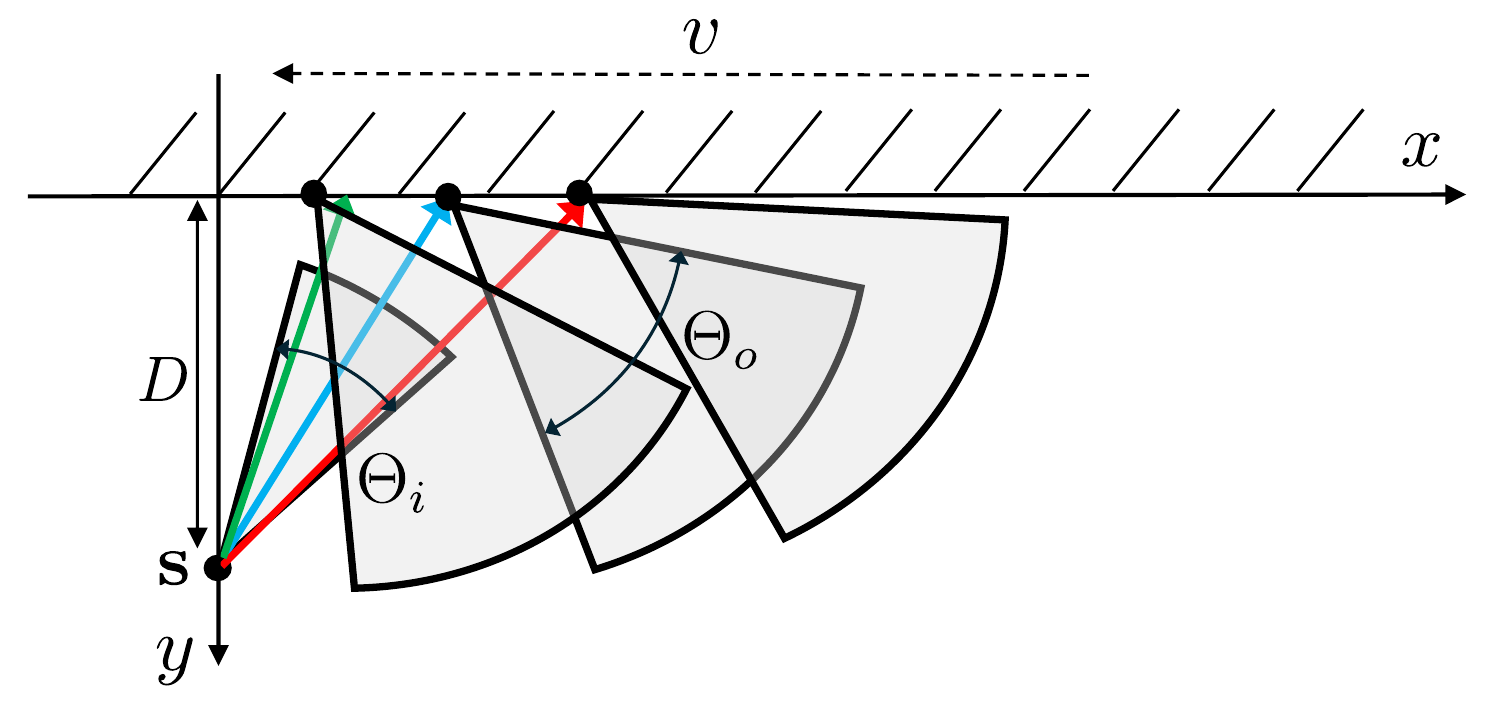}}
    \caption{Sketch of a system with a moving source and an anomalous reflection plane: (a) the source illuminates the plane with a varying $\theta_i$, and the plane reflects according to the specific incident point $\mathbf{p}$; (b) the system (a) is equivalent to a fixed source and multiple snapshots over double sweeping with the angular spans $\Theta_i$ and $\Theta_o$.}
    \label{fig:varying_inc_varying_refl}
\end{figure}

Consider now a slightly different setup, illustrated in Fig.~\ref{fig:varying_inc_varying_refl}, where the moving source emits periodic pulses while varying the incidence angle $\theta_i$ at each pulse, i.e., $\theta_i(\tau)$. The latter incidence angle spans another angular set $\Theta_i$.
The reflection plane implements a space-varying reflection angle $\theta_o(x) \equiv \theta_o(\tau)\in \Theta_o$. The Rx data consists of a data cube formed by triplets $(\theta_i,\theta_o,\tau)$. By sweeping both transmitted and reflected angles, $\theta_i(\tau)\in \Theta_i$, $\theta_o(\tau)\in \Theta_o$, it is possible to obtain both sensing coverage \textit{and} improved resolution, as described in Section \ref{sec:system_design}. In the following, we detail this novel concept of \textit{stroboscopic sensing}, where a moving source can observe (sense) a moving target through multiple snapshots as if it were static.


\begin{figure}
    \centering
    \subfloat[][Imaging with a mirror]{\includegraphics[width=0.75\columnwidth]{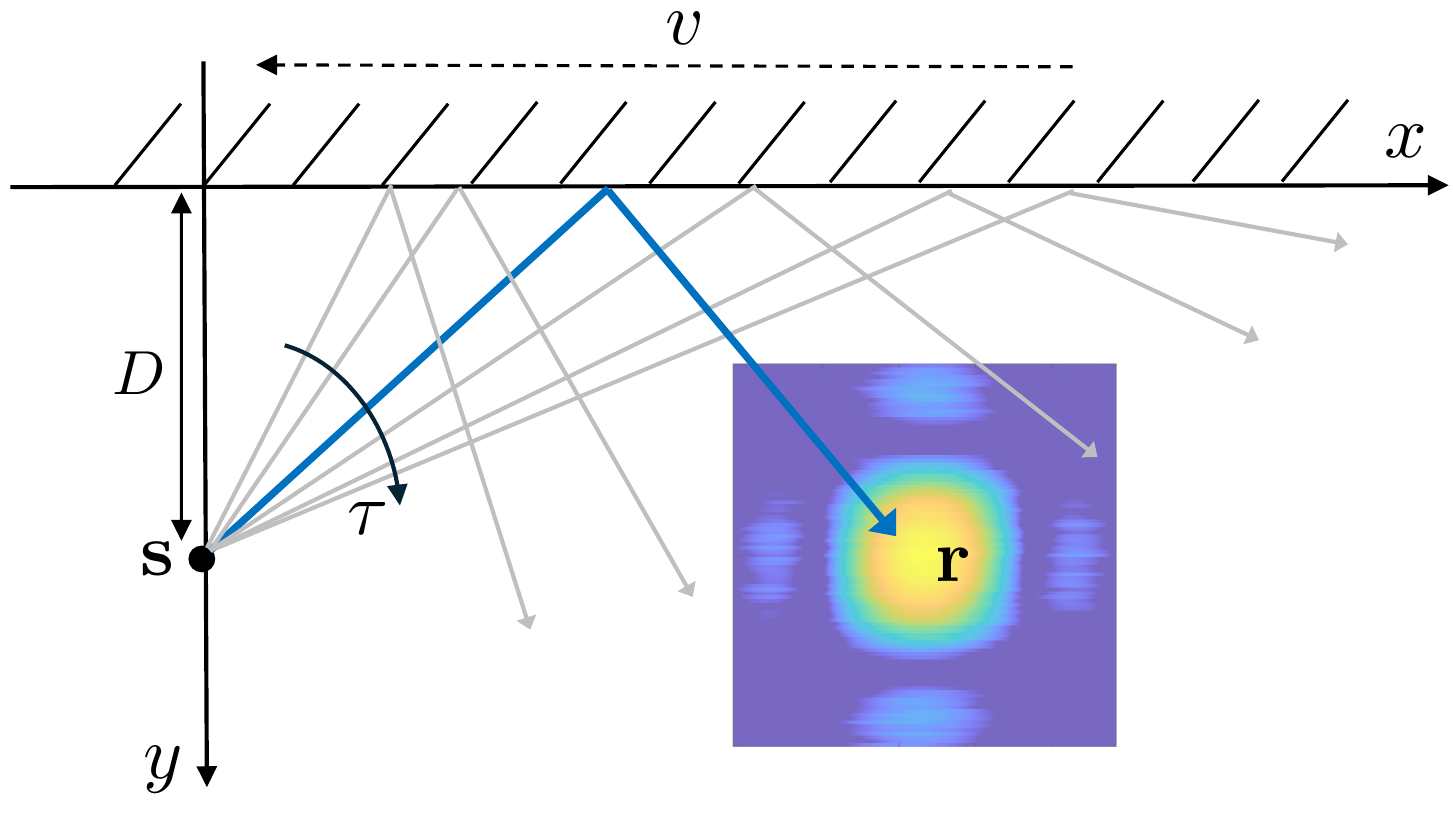}} \\ \vspace{-0.25cm}
    \subfloat[][Imaging with the proposed stroboscopic system]{\includegraphics[width=0.75\columnwidth]{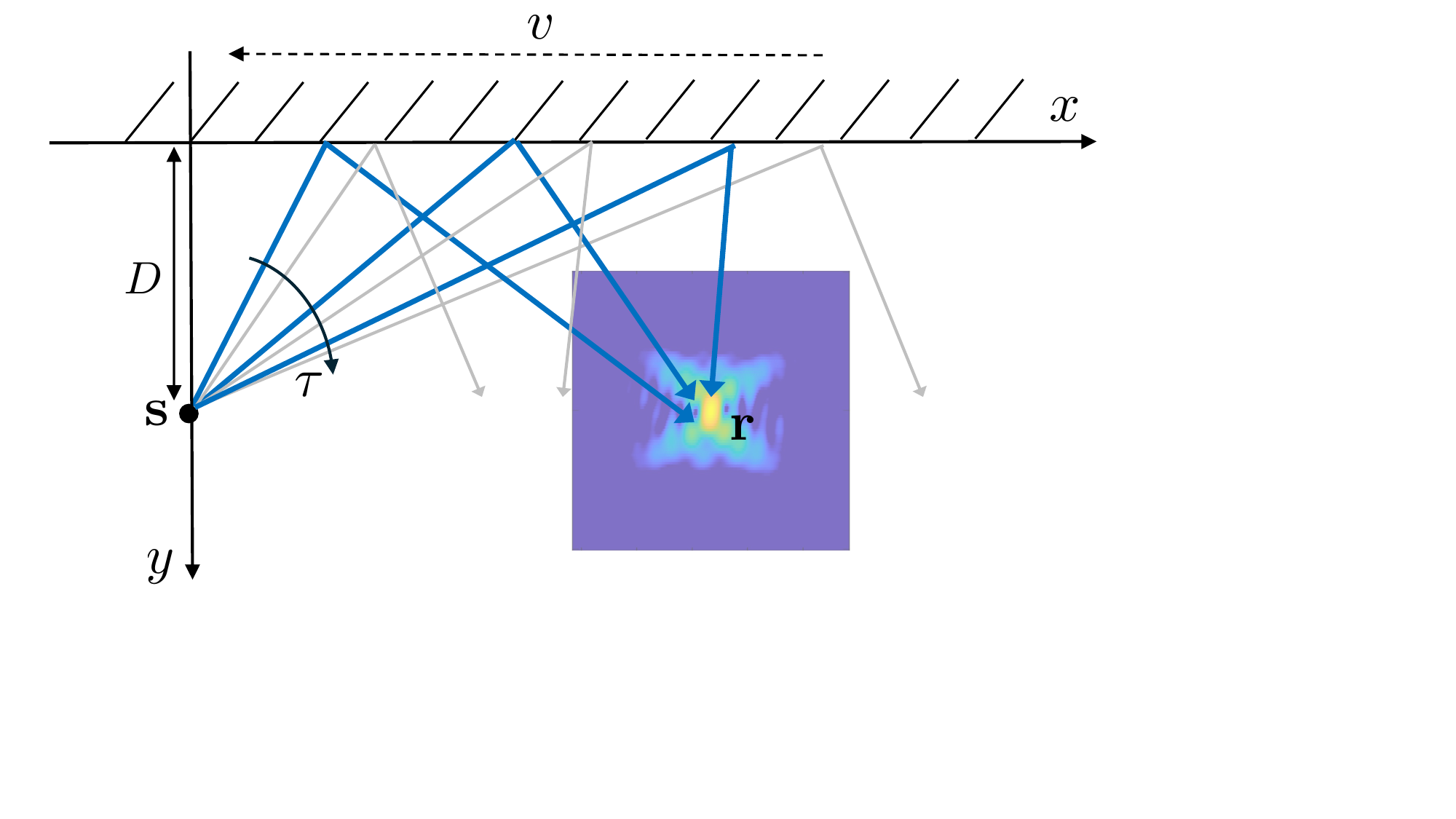}}
    \caption{Example of point target imaging with (a) a mirror and (b) the proposed stroboscopic system with $P > 1$ viewpoints. By sweeping the Tx beam over $\Theta_i$ and using a reflection plane as for \eqref{eq:angle_periodic_design}, it is possible to increase the image resolution by multiple view-points (blue propagation lines). }
    \label{fig:strobo_spiegazione}
\end{figure}

\subsection{Stroboscopic Sensing: the Key Idea}

In this paper, the aforementioned stroboscopic system is exploited for creating the image of a desired region of interest (ROI) in NLOS, that moves at the same speed of the source $v$, in the same direction. For simplicity, we herein consider a rectangular ROI of size $\Delta_x$ and $\Delta_y$ along $x$ and $y$. In this reference system, the source in $\mathbf{s} = (0,D)$, is emitting multiple pulses within $\Theta_i$, for a total observation time $T_\text{obs}=|\Theta_i| \Delta \tau$. A target in the ROI is located in $\mathbf{r} = (r_x,r_y)$. The reflection angle experienced at snapshot time $\tau=\ell \Delta \tau$ is: 
\begin{equation}\label{eq:angle_periodic_design_varyingincidence}
\begin{split}
        \theta_o(\tau) = \theta_i(\tau) & + \Delta \theta(x)\bigg\lvert_{x=  v \tau + D \tan \theta_i(\tau)}.
\end{split}
\end{equation}
The reflection plane implements a periodic anomalous pattern such that to reflect over the set of angles $\Theta_o$ through the space-varying angular difference:
\begin{equation}\label{eq:angle_periodic_design}
   \Delta \theta(x) = \overline{\theta}_o - \overline{\theta}_i +  \frac{\Delta\theta_{o,\text{obs}}}{2}\cos\left(\frac{2 \pi}{\Lambda} x \right)
\end{equation}
where $\Lambda$ is the spatial periodicity of $\theta_o(x)$ around a pair of incidence and reflection angles $\overline{\theta}_i$ and $\overline{\theta}_o$, whose values are design parameters detailed in \ref{sec:system_design}.
A sinusoidal function is not the only possibility for $\Delta \theta(x)$, but any periodic function provides the same angular coverage that sweeps $\theta_o(x)$ while moving. 

Now, the problem is to illuminate the target in $\mathbf{r} = [r_x, r_y]$. For a given arbitrary incidence angle $\theta_i$, the reflection angle that guarantees to intercept the target is
\begin{equation}\label{eq:angle_reflection_target}
    \theta_o(\theta_i|\mathbf{s},\mathbf{r}) = \arctan\left(\frac{r_x - D \tan \theta_i}{r_y}\right).
\end{equation}
By plugging $\Delta \theta(x)$ of \eqref{eq:angle_periodic_design} in \eqref{eq:angle_periodic_design_varyingincidence}, we can equate the result with \eqref{eq:angle_reflection_target}. The target will be effectively illuminated only in those time instants obtained by finding the $P$ roots of $\theta_o(\tau) =  \theta_o(\theta_i|\mathbf{s},\mathbf{r})$, $\tau\in[0,T_{\mathrm{obs}}]$. If $P=1$, the target is illuminated from a single point over the plane, thus the resolution of the final image is approximately dictated by source's aperture. Differently, if $P>1$  it follows that  $\theta_o(\tau_1)\neq\theta_o(\tau_2)\neq...\neq\theta_o(\tau_P)$, the target is illuminated from multiple view-points, and the resolution of the radar image may increase, as detailed in Appendix~\ref{sec:resolution}. Fig. \ref{fig:strobo_spiegazione} illustrates the advantage of using the stroboscopic system w.r.t. a mirror. By sweeping the Tx beam over $\Theta_i$ and a properly configured reflection plane, the image of a target improves thanks to multiple view-points, as illustrated in Fig. \ref{fig:strobo_spiegazione}b. In some cases, the resolution enhances along both $x$ and $y$ thanks to the near-field effect (see Section \ref{sec:system_design} for further details).

\begin{figure}[t]
    \centering
    \includegraphics[width=0.8\columnwidth]{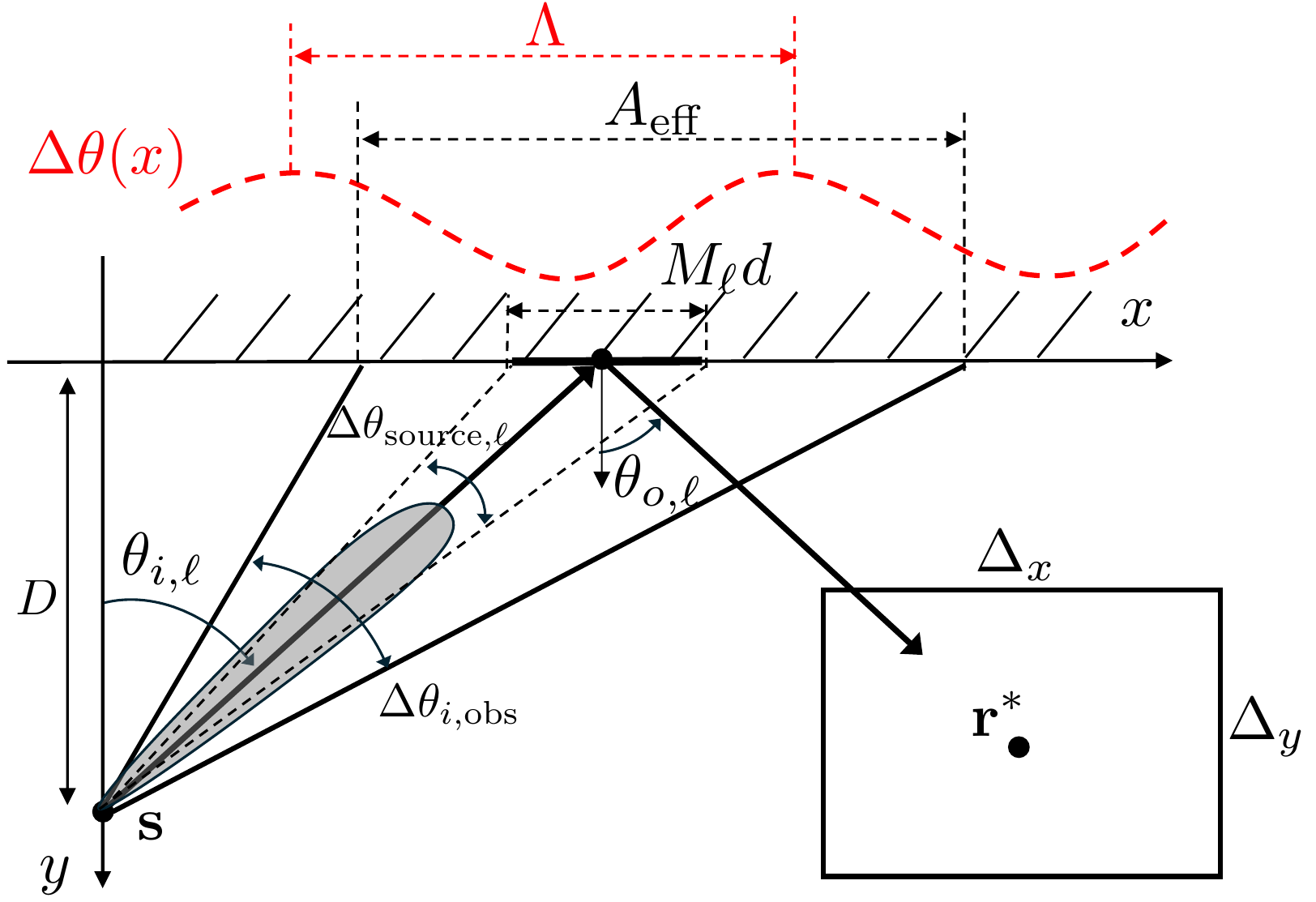}
    \caption{Sketch of the considered geometry.}
    \label{fig:system_model}
\end{figure}

\section{System Model}\label{sec:system_model}

Making reference to the system illustrated in Section \ref{sect:multiview}, we consider a moving source whose positions along the slow time interval $\tau = \ell \Delta \tau \in[0,T_\mathrm{obs}]$ are $\mathbf{s}(s_x + \ell \Delta \tau) \equiv \mathbf{s}_\ell = (s_x(0) + \ell v\Delta \tau, D)$, $\ell=1,..,|\Theta_i|$ slow time samples. Similarly, the ROI of size $\Delta_x \times \Delta_y$ moves in the same direction and at the same speed $v$, as if in a vehicular setup where ROI propagation is blocked. ROI center is located in $\mathbf{r}^*_\ell = (r^*_x(0)+\ell v\Delta \tau, r^*_y)$. 
The source is equipped with a sensing platform ---either a radar or an ISAC transceiver---emitting the passband signal: 
\begin{equation}\label{eq:TX_signal}
    g_\text{RF}(t) = g_\text{BB}(t) e^{j2\pi f_0 t}
\end{equation}
where $g_\text{BB}(t)$ is the base-band range-compressed pulse with bandwidth $B$. Tx signal \eqref{eq:TX_signal} is emitted periodically with interval $\Delta \tau$. The source has a physical aperture $A$ along $x$, allowing beamforming of \eqref{eq:TX_signal} along direction $\theta_i(\ell \Delta \tau) \equiv \theta_{i,\ell}$ with a beamwidth
\begin{equation}\label{eq:source_bw}
    \Delta\theta_{\text{source},\ell} \simeq \frac{\lambda_0}{A \cos \theta_{i,\ell}}, 
\end{equation}
where $\lambda_0 = c/f_0$ is the carrier wavelength. The reflection plane is made of a planar metasurface, composed by $N$ meta-atoms. In global coordinates, the position of the $n$th meta-atom is $\mathbf{p}_n = (nd, 0)$,  $n=0,...,N-1$, where $d$ is the inter-element distance. At the $\ell$th time instant, the source illuminates the portion of metasurface composed of:
\begin{equation}
\label{eq:Nrad}
    M_\ell \approx \frac{D \Delta\theta_{\text{source},\ell}}{d \cos \theta_{i,\ell} \sin \theta_{i,\ell}}
\end{equation}
elements, within the meta-atoms set 
\begin{equation}
    \mathcal{M}_\ell = \left\{m \, \bigg \lvert \, m= n_{0,\ell} -\frac{M_\ell}{2}, ..., n_{0,\ell} + \frac{M_\ell}{2} -1\right\}
\end{equation}
where 
\begin{equation}
    n_{0,\ell} = \frac{\ell v \Delta \tau + D \tan \theta_{i,\ell}}{d}
\end{equation}
is the element index corresponding to the peak of the radiation beam, located in position $(\ell v \Delta \tau + D \tan \theta_{i,\ell},0)$. The union of sets $\mathcal{M}_\ell$, $\ell=1,...,L$, along the observation time $T_\mathrm{obs}$, forms the so-called \textit{effective aperture} $A_\text{eff}$ (see Fig. \ref{fig:system_model}), that dictates the resolution of the resulting image of the target, as detailed in Appendix \ref{sec:resolution}. The aperture $A_\text{eff}$ can be approximated for narrow beams as
\begin{equation}\label{eq:Aeff}
    A_\text{eff} \simeq D \left[ \tan\left(\overline{\theta}_i \hspace{-0.05cm} + \hspace{-0.05cm} \frac{\Delta \theta_{i,\text{obs}}}{2}\right) - \tan\left(\overline{\theta}_i \hspace{-0.05cm}-\hspace{-0.05cm} \frac{\Delta \theta_{i,\text{obs}}}{2}\right)\right].
\end{equation}
for a certain direction $\overline{\theta_i}$.
The system model and the notation are illustrated in Fig. \ref{fig:system_model}.

The model for the Rx signal at the source ($\ell$th slow-time sample) is reported in \eqref{eq:Rx_signal_generic_narrowbeam},
\begin{figure*}[t]
\begin{equation}\label{eq:Rx_signal_generic_narrowbeam}
\begin{split}
       y_\ell(t) = \rho_\ell \, e^{j \psi} \, g\left(t- \frac{2\left[D_{i,\ell} + D_{o,\ell}\right]}{c}\right)& e^{-j \frac{4 \pi}{\lambda_0} \left[D_{i,\ell} + D_{o,\ell}\right]}  \sum_{n\in \mathcal{M}_\ell} 
        \sum_{n'\in \mathcal{M}_\ell} 
        e^{j \phi_n}e^{j\phi_{n'}}  e^{-j \frac{2\pi d}{\lambda_0}(n+n') \left[ \sin\theta_{i,\ell} -  \sin\theta_{o,\ell}\right]}
         + w_\ell(t)
\end{split}
\end{equation}\hrulefill
\end{figure*}
where \textit{(i)} $\rho_\ell$ is the path-loss considering the double reflection off the single atom of the metasurface as well as the target's radar cross section (RCS), \textit{(ii)} $\psi$ is the scattering phase of the target, \textit{(iii)} $g(t) = g_\text{BB}(t) * g^*_\text{BB}(-t)$ is the pulse after the matched filtering, \textit{(iv)} $D_{i,\ell} = \| \mathbf{p}_{0,\ell} - \mathbf{s}_\ell \|$ and $D_{o,\ell} = \| \mathbf{r}_k - \mathbf{p}_{0,\ell} \|$ are the source-metasurface and target-metasurface distances, respectively (considering the beam center), \textit{(v)} $\phi_n$ is the reflection phase applied at the $n$th element of the metasurface and \textit{(vi)} $w_\ell(t)$ is the additive white Gaussian noise with power $\sigma_w^2$. The path-loss $\rho_\ell$ follows from the radar equation applied to a double-bounce off a single meta-atom:
\begin{equation}
    \rho_\ell = \sqrt{\frac{B T_s\lambda_0^6 \,\eta^2_{\text{source}}(\theta_{i,\ell})\, \eta^2_{\text{atom}}(\theta_{i,\ell})\,  \eta^2_{\text{atom}}(\theta_{o,\ell})\, \sigma_{\mathbf{r}}}{(4 \pi)^7 D_{i,\ell}^4 D_{o,\ell}^4}}
\end{equation}
where $T_s \leq \Delta \tau$ is the duration of the Tx signal (before matched filtering), $\eta_{\text{source}}(\theta_{i,\ell})$ is the beamforming gain provided by the source, $\eta_{\text{atom}}(\cdot)$ is the radiation power pattern of the single meta-atom, evaluated for incidence ($\theta_{i,\ell}$) and reflection ($\theta_{o,\ell}$) angles, while $\sigma_{\mathbf{r}}$ is the RCS of the target, that we assume to be isotropic. In this system, $\eta_{\text{source}} \gg \eta_{\text{atom}}$.  
Rx signal model \eqref{eq:Rx_signal_generic_narrowbeam} is extended to the multi-target case in a straightforward manner (see Section \ref{sec:results}). 

We assume herein that both \textit{far-field} and \textit{spatial narrowband} assumptions hold. The former assumes that the beamwidth $\Delta\theta_{\text{source},\ell}$ is narrow enough to have a planar wavefront impinging the reflection plane over the $M_\ell$ meta-atoms. The spatial narrowband condition, instead, holds for 
\begin{equation}\label{eq:narrowband_condition}
    \frac{1}{B} \gg M_\ell \frac{d}{c} \mathrm{max}\left\{\sin \theta_{i,\ell}, \sin \theta_{o,\ell}\right\}, \,\,\,\,\, \forall \ell,
\end{equation}
meaning that the residual propagation delay across the effective array is much less than the pulse duration $1/B$ (i.e., the bandwidth of the signal is small enough to neglect any beam squinting effects on the illuminated portion of metasurface). Notice that \eqref{eq:narrowband_condition} is function of the Tx beamwidth \eqref{eq:source_bw} (via $M_\ell$ and thus the incidence angle $\theta_{i,\ell}$) as well as of the reflection angle $\theta_{o,\ell}$. While the far-field condition usually applies for beams $\Delta\theta_{\text{source}}$ in the order of some degrees and distances of few to tens of meters, the spatially narrowband operation is strongly dependent on the employed bandwidth. For instance, for $\theta_i=60$ deg, $D_i=10$ m, we obtain that $B < 100$ MHz for $\Delta\theta_{\text{source}} = 10$ deg, and $B<1$ GHz for $\Delta\theta_{\text{source}} = 1$ deg, the latter being the current limit for off-the-shelf automotive radars in terms of angular resolution~\cite{TI_ref_MMWCAS}.
The far-field and spatial narrowband assumptions does not limit the generality of this work, since the near-field operation implies dealing with multiple incidence angles $\theta_{i,\ell}$ at the same slow-time instant $\ell$, while the narrowband one typically applies for bandwidths in the order of tens of MHz. However, the Rx signal model in \eqref{eq:narrowband_condition} simplifies the conveyance of the main concepts ruling the design of the proposed stroboscopic sensing system.

\section{Image Formation and Angular Sampling Limit}\label{sec:image_formation}
\begin{figure}[!t]
    \centering
    \subfloat[][]{\includegraphics[width=0.7\columnwidth]{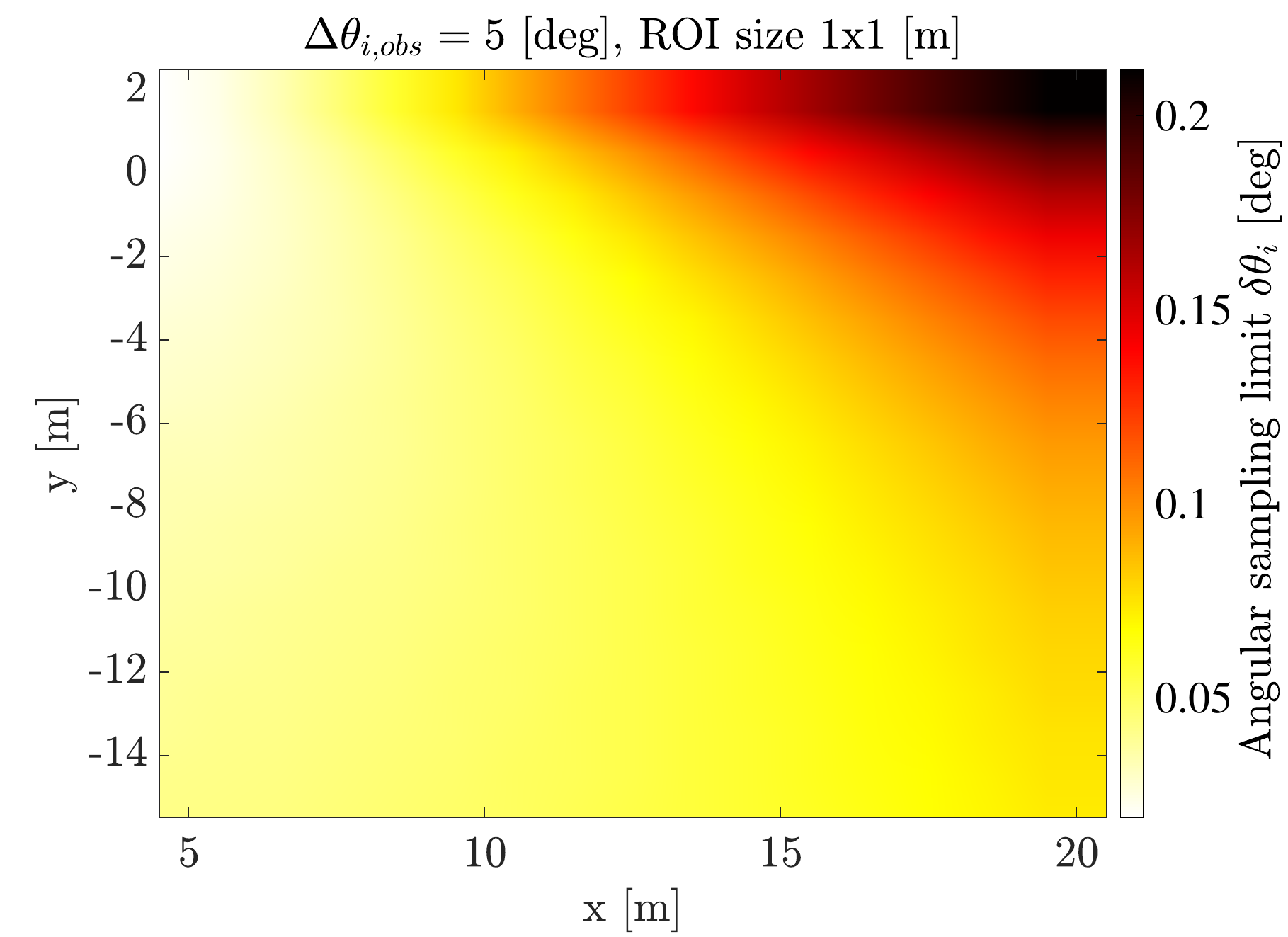}\label{subfig:5deg}}\\ \vspace{-0.25cm}
    \subfloat[][]{\includegraphics[width=0.7\columnwidth]{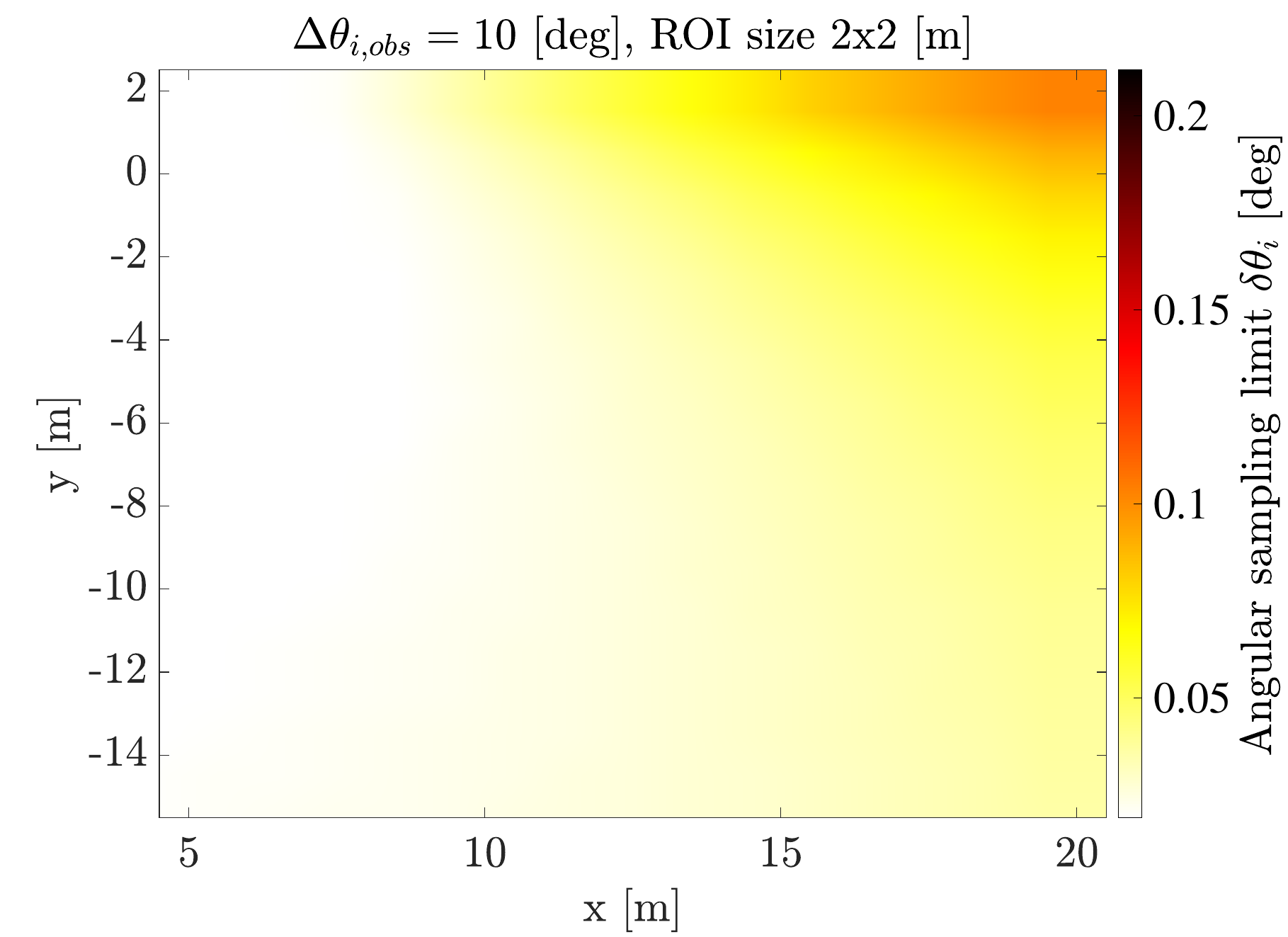}\label{subfig:10deg}}
    \caption{Angular Sampling limit $\delta \theta_i$ as function of the position of the ROI $\mathbf{r}^*$, for $f_0=77$ GHz, $D=5$ m, $\overline{\theta}_i=30$ deg: (a) $\Delta \theta_{i,\mathrm{obs}} = 5$ deg, $1 \times 1$ m$^2$ ROI, (b) $\Delta \theta_{i,\mathrm{obs}} = 10$ deg, $2 \times 2$ m$^2$ ROI.  }\label{fig:sampling_cond}
\end{figure}
The image of the scatters within the ROI is formed by matched filtering over fast and slow time, i.e., 2D back-projection (BP). The BP is the correlation between the Rx signal and the noiseless model of the Rx signal from an arbitrary location $\mathbf{x}=(x,y)$ within the ROI, providing the likelihood of the targets' position in $\mathbf{x}$ (see \cite{5739256} for further details). The image of a generic ROI composed of a target in $\mathbf{r}$ is therefore synthesized as follows:
\begin{equation}\label{eq:BP}
\begin{split}
        I(\mathbf{x}) &= \sum_\ell y_\ell\left(t \hspace{-0.05cm} = \hspace{-0.05cm} \frac{2\left[D_{i,\ell} \hspace{-0.05cm}+ \hspace{-0.05cm}D_{o,\ell}(\mathbf{x})\right]}{c}\right) \hspace{-0.05cm}e^{j \varphi(\theta_{i,\ell} | \mathbf{x})} \\
        & = \alpha \,e^{j \psi} H(\mathbf{x}-\mathbf{r}) + Z(\mathbf{x})
\end{split}
\end{equation}
where $\varphi (\theta_{i,\ell} | \mathbf{x}) = \frac{4\pi}{\lambda} [D_{i,\ell} + D_{o,\ell}(\mathbf{x})]$ is the propagation phase, $\mathbf{x}$ is a pixel of the grid representing the ROI and $D_{o,\ell}(\mathbf{x}) = \| \mathbf{x} - \mathbf{p}_{0,\ell}\| $ is the distance from the interception point on the plane and the considered pixel. The result of the BP is the a scaled and shifted replicas of the \textit{spatial ambiguity function} $H(\mathbf{x})$, i.e., the image of a point target in the considered setup. Noise in the spatial domain is denoted with $Z(\mathbf{x})$, and it is generally spatially correlated. BP follows three cascaded processing steps on the Rx signal $y_\ell(t)$ in \eqref{eq:Rx_signal_generic_narrowbeam}: \textit{(i)} evaluation at the time of flight corresponding to a scattering from pixel $\mathbf{x}$ via the reflection off the plane (possibly with a suitable interpolation for sampled signals), \textit{(ii)} compensation for the propagation phase by $e^{j \varphi (\theta_{i,\ell}|\mathbf{x})}$ and \textit{(iii)} summation over all the slow-time snapshots $\tau= \ell \Delta \tau$.

Imaging \eqref{eq:BP} is achieved by coherently combining the Rx echoes from the targets in the ROI for each Tx angle $\theta_{i,\ell}$ within the angular set:
\begin{equation}\label{eq:incidence_codebook}
    \Theta_i = \left\{\overline{\theta}_i -\frac{\Delta \theta_{i,\text{obs}}}{2} : \delta \theta_i : \overline{\theta}_i + \frac{\Delta \theta_{o,\text{obs}}}{2}\right\}
\end{equation}
where $\delta \theta_i$ is the angular sampling interval and $\overline{\theta_i}$ is a reference incident angle. We assume that $\Delta \theta_{i,\text{obs}}$ is larger than (or much larger) than the source beamwidth $\Delta \theta_\text{source}$\footnote{Since the source emits a pulse at each snapshot on a pre-defined set of angles, the integration over the employed Tx beams is equivalent to the integration over the slow time (snapshots).}. The angular sampling interval $\delta \theta_i$ to be used when progressively illuminating the reflection plane is of fundamental importance to avoid grating lobes in the resulting image of a given target within the ROI. In other words, the spatial ambiguity function $H(\mathbf{x})$ shall exhibit a single global maximum within the ROI. To avoid spatial aliasing, we analyze the variation of the instantaneous spatial frequency (usually defined as the rate at which the propagation phase changes as the wave propagates through space) w.r.t. an infinitesimal variation of the Tx angle $\theta_i$. Recalling the coordinate transformation where the source is in $\mathbf{s}=(0,D)$ and the target is in $\mathbf{r} = (r_x , r_y)$, the phase derivative is
\begin{equation}\label{eq:ph_derivative}
\begin{split}
      \frac{\mathrm{d}\varphi (\theta_i|\mathbf{r})}{\mathrm{d}\theta_i } & = \frac{\mathrm{d}}{\mathrm{d} \theta_i} \hspace{-0.1cm} \left[ \frac{4 \pi}{\lambda_0} (D_i(\theta_i) + D_o(\theta_i|\mathbf{r}))\right] \\ 
      & = \frac{4 \pi D}{\lambda_0 \cos^2\theta_i} \left( \sin\theta_i -\frac{r_x \hspace{-0.1cm}- \hspace{-0.1cm}D \tan \theta_i}{\sqrt{r_y^2 + (r_x \hspace{-0.1cm}-\hspace{-0.1cm} D \tan \theta_i)^2}}\right) \\
      & = \frac{4 \pi D}{\lambda_0 \cos^2\theta_i} \left( \sin\theta_i - \sin\theta_o(\theta_i|\mathbf{s},\mathbf{r}) \right).
\end{split}
\end{equation}
It is a monotonically increasing function with $\theta_i$, since the path length increases with $\theta_i$ as well. The sampling interval is therefore ruled by the maximum difference between the phase derivatives within $\Delta \theta_{i,\mathrm{obs}}$, namely:
\begin{equation}\label{eq:deltathetai}
    \delta \theta_i \leq \frac{ \pi}{\bigg \lvert \underset{\mathbf{r}}{\max} \left\{ \frac{\mathrm{d}\varphi (\theta_i|\mathbf{r})}{\mathrm{d}\theta_i } \big \rvert_{\frac{\Delta \theta_{i,\mathrm{obs}}}{2}}\right\} - \underset{\mathbf{r}}{\min} \left\{  \frac{\mathrm{d}\varphi (\theta_i|\mathbf{r})}{\mathrm{d}\theta_i } \big \rvert_{-\frac{\Delta \theta_{i,\mathrm{obs}}}{2}}\right\} \bigg \rvert}
\end{equation}
where the dependence on the target's position $\mathbf{r}$ is made explicit for convenience.
In practice, testing the corners of the rectangular ROI guarantees that each target is imaged without aliasing. 
An example of the sampling condition $\delta \theta_i$ is shown in Fig. \ref{fig:sampling_cond}, for $f_0=77$ GHz, $D=5$ m, $\overline{\theta}_i=30$ deg, reported as a function of the ROI center $\mathbf{r}^*$ and $\Delta \theta_{i,\mathrm{obs}}$. Increasing the size of the ROI $\Delta_x \times \Delta_y$ and $\Delta \theta_{i,\mathrm{obs}}$ decreases the required angular sampling time. To relax the angular sampling \eqref{eq:deltathetai} and reduce the overall observation time $T_\text{obs}$, the only option is to limit the angular observation interval $\Delta \theta_{i,\mathrm{obs}}$, thus the overall resolution of the imaging system (as discussed in Section \ref{sec:system_design}). This trade-off characterizes the proposed system and it is dependent on the application domain. The effective image resolution depends on $\Delta \theta_{i,\mathrm{obs}}$ via $A_\text{eff}$ in \eqref{eq:Aeff} as well on other design considerations discussed in the following Section \ref{sec:system_design}.

\section{Design of the System}\label{sec:system_design}
\begin{figure}[b]
    \centering
    \includegraphics[width=0.9\columnwidth]{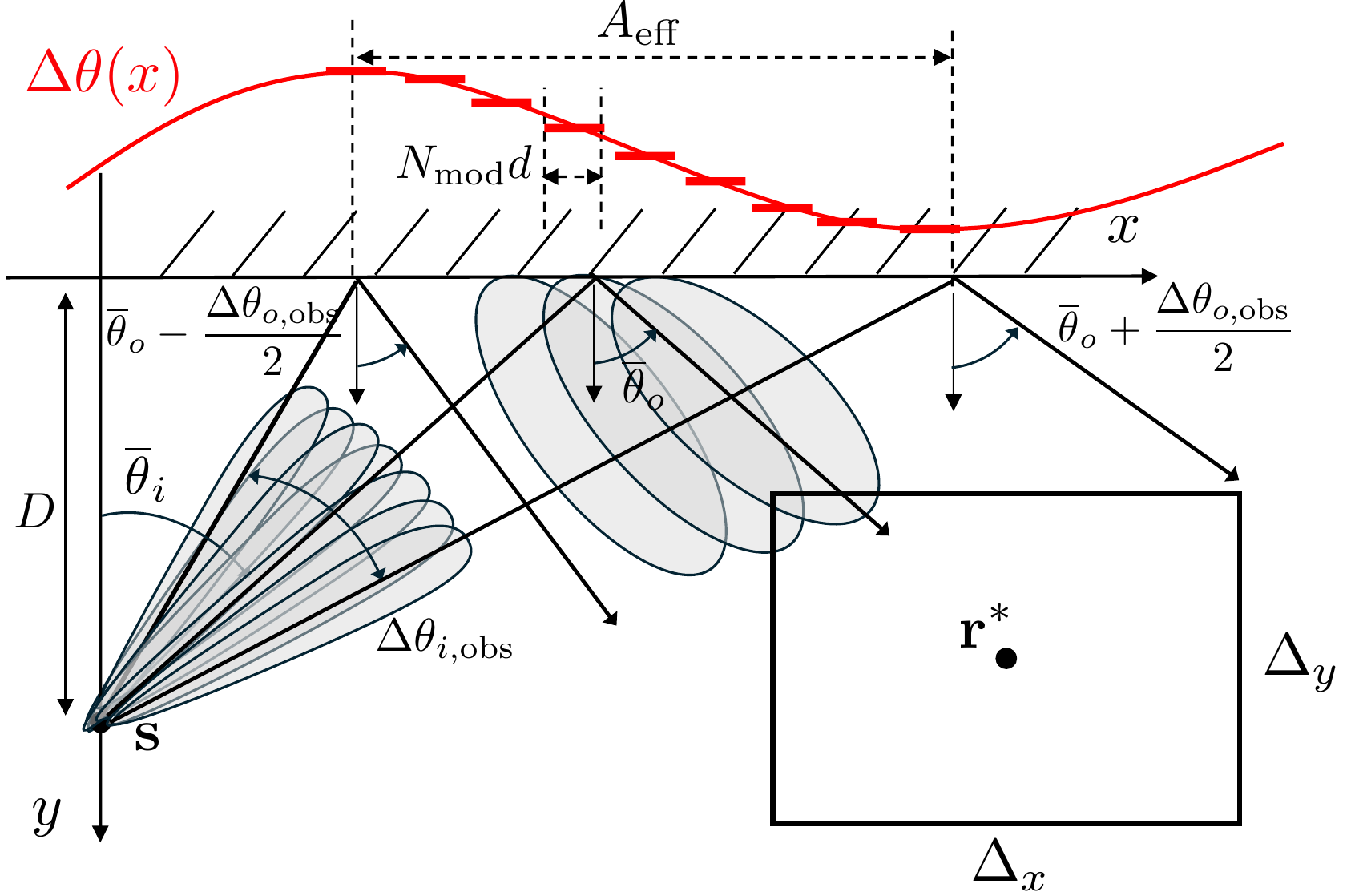}
    \caption{Sketch of the designed system. }
    \label{fig:system_design_sketch}
\end{figure}

The anomalous reflection plane considered in the paper is implemented with a sequence of SP-EMS (here also referred to as \textit{modules}). For sake of reasoning, we opt for such modular implementation of the SP-EMS to replace the continuous phase gradients with a finite set of reflection angles. A modular implementation allows approximating a near-field design involving the whole reflection plane with a module-by-module far-field design. Remarkably, the performance of such a system is comparable with the one designed for near-field operation, as detailed in~\cite{10541333}.
From anomalous reflection angles escribed in \eqref{eq:angle_periodic_design} one needs to use a finite set of angles
\begin{equation}\label{eq:angle_periodic_design_quant}
\begin{split}
        \Delta\theta_{\text{mod}}(x) = \mathcal{Q}_{|\Theta_o|}\left[ (\overline{\theta}_o-\overline{\theta}_i) + \frac{\Delta\theta_{o,\text{obs}}}{2}\cos\left(\frac{2 \pi}{\Lambda} x\right)\right]
\end{split}
\end{equation}
where $\mathcal{Q}_{|\Theta_o|}\left[\cdot\right]$ is the quantization function over the set of reflection angles $\Theta_o$. The average number of elements per module is 
\begin{equation}
    N_\text{mod} \simeq \frac{\Lambda}{2 d |\Theta_o|}
\end{equation}
where factor $2$ indicates that the reflection angles $\Theta_o$ are fully explored in half cycle of the spatial sinusoidal function, for $x\in[0,\Lambda/2]$.
The required phase gradient to apply at the SP-EMS to implement the discrete angle set \eqref{eq:angle_periodic_design_quant} is therefore:
\begin{equation}
\begin{split}
    \phi(x) =  \frac{2 \pi}{\lambda_0} x \left[\sin \overline{\theta}_i - \sin \left(\overline{\theta}_i + \Delta\theta_{\text{mod}}(x)\right)\right] \bigg\rvert_{x=nd}.
\end{split}
\end{equation}
The SP-EMS is designed according to the following parameters
\begin{align}
    \overline{\theta}_o & = \theta_o(\overline{\theta_i}|\mathbf{s},\mathbf{r}^*)  \label{eq:theta_o_center}\\
    \Delta\theta_{o,\text{obs}} &= \theta_o(\overline{\theta_i}|\mathbf{s},\mathbf{r}^+) - \theta_o(\overline{\theta_i}|\mathbf{s},\mathbf{r}^-)\label{eq:theta_o_max}
\end{align}
where $\theta_o(\overline{\theta_i}|\mathbf{s},\mathbf{r}^*)$, $\theta_o(\overline{\theta_i}|\mathbf{s},\mathbf{r}^+)$ and $\theta_o(\overline{\theta_i}|\mathbf{s},\mathbf{r}^-)$ are the reflection angles associated to the center of the ROI $\mathbf{r}^*$ and the two corner points $\mathbf{r}^+ = \mathbf{r}^* + (\Delta_x/2,\Delta_y/2)$ and $\mathbf{r}^- = \mathbf{r}^* - (\Delta_x/2,\Delta_y/2)$, respectively, obtained as \eqref{eq:angle_reflection_target}. The design according to \eqref{eq:theta_o_center} and \eqref{eq:theta_o_max} allows spanning the ROI during the reflection sweeping.

\subsection{Selection of the Reflection Codebook $\Theta_o$}\label{subsec:reflection_sampling}

\begin{figure}[!t]
   \centering
   \subfloat[][$N_\text{mod} > N^{\max}_\text{mod}$]{\includegraphics[width=0.5\columnwidth]{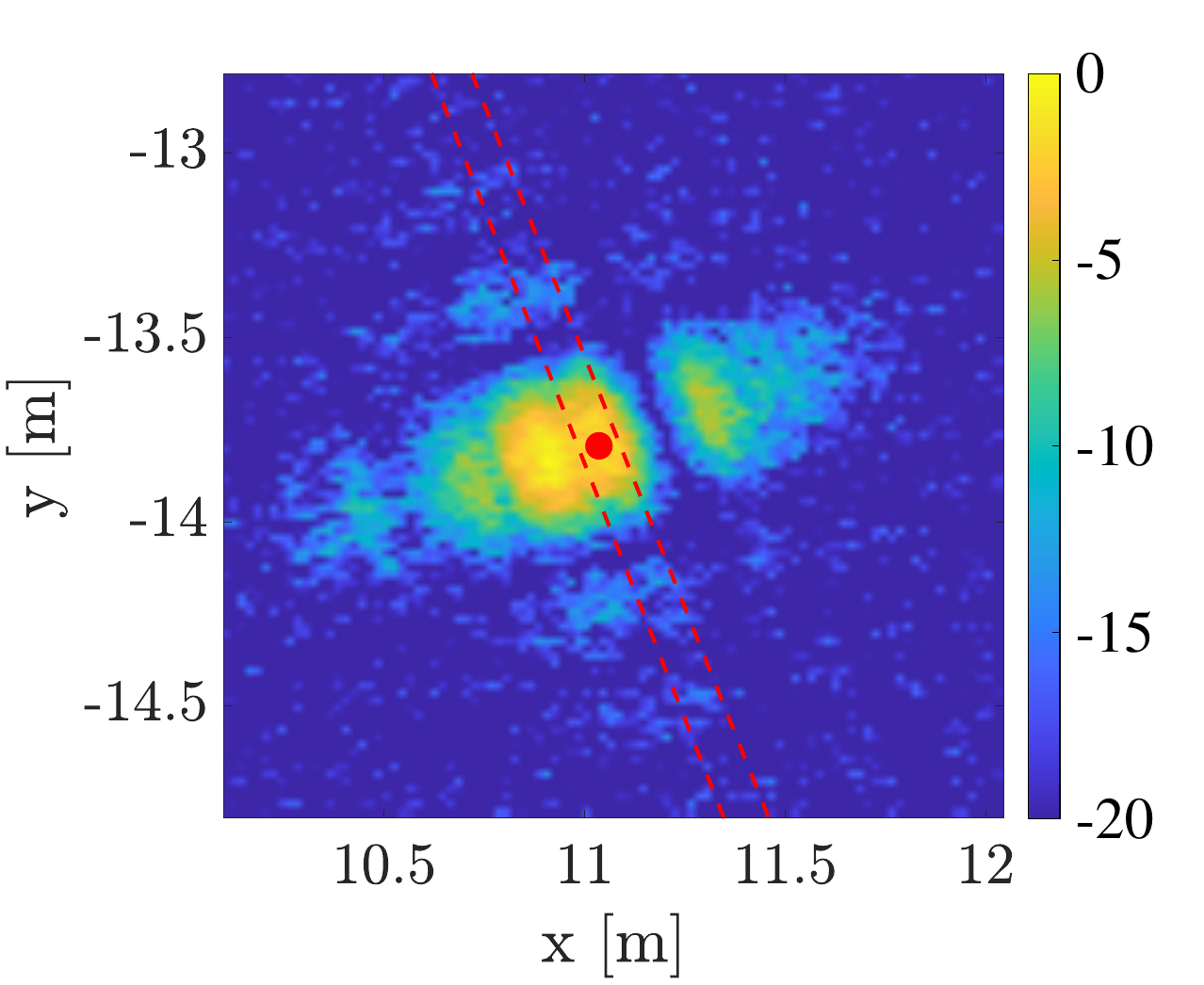}\label{subfig:large_modules}}
   \subfloat[][$N_\text{mod} < N^{\max}_\text{mod}$]{\includegraphics[width=0.5\columnwidth]{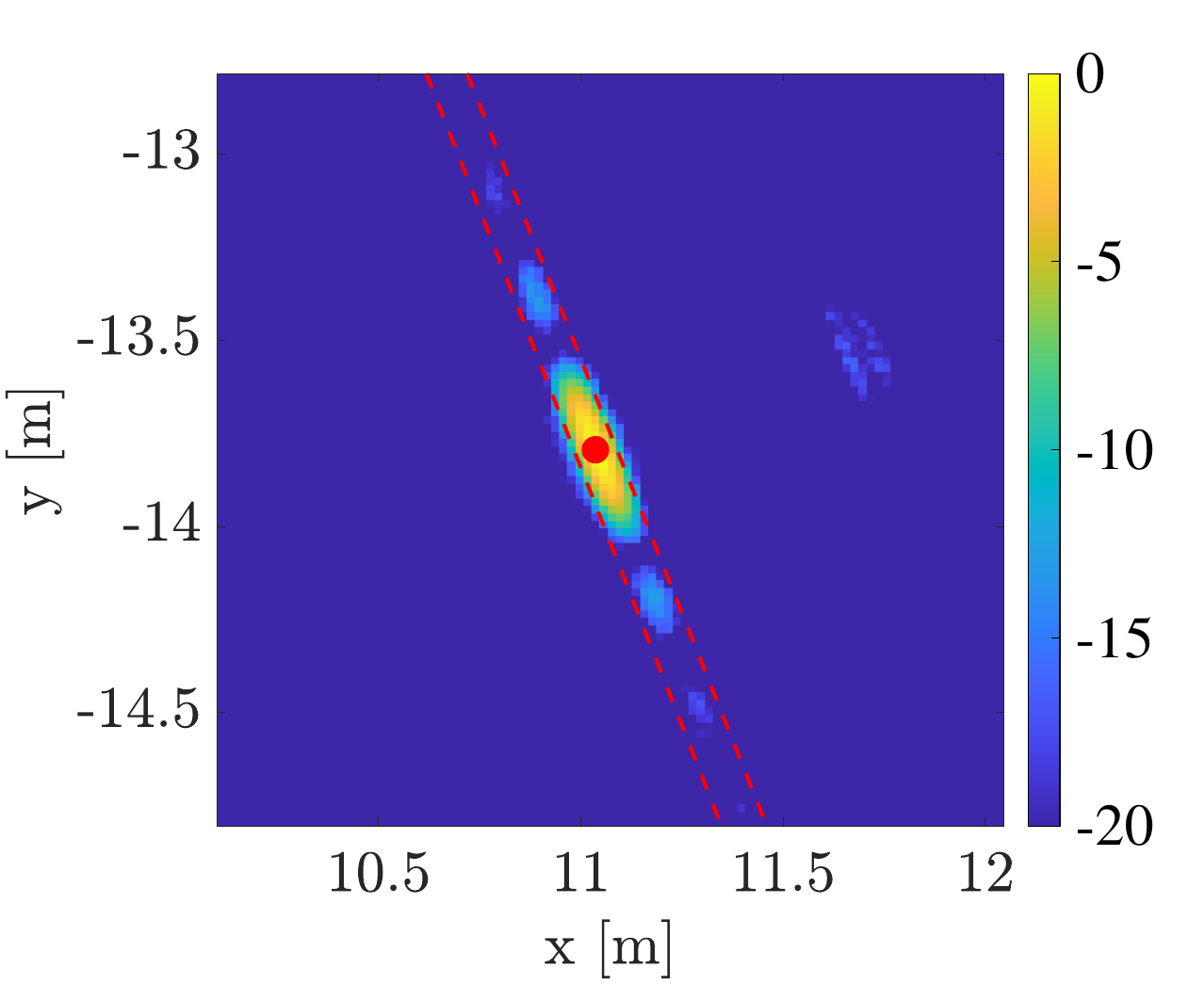}\label{subfig:small_modules}}
   \caption{Image example of a point target illuminated by (a) few large modules, thus $N_\text{mod} > N^{\max}_\text{mod}$ and (b) many small modules, thus $N_\text{mod} < N^{\max}_\text{mod}$. The image on the left is affected by a severe loss of resolution compared to the one on the right, since only one module illuminates the target. Moreover, the peak of the image (a) is not centered in the true target position. }
    \label{fig:module_size}
\end{figure}

The reflection codebook implemented by the SP-EMS is: 
\begin{equation}\label{eq:reflection_codebook}
   \Theta_o = \left\{ \overline{\theta}_o - \frac{\Delta \theta_{o,\text{obs}}}{2} : \delta \theta_o :  \overline{\theta}_o + \frac{\Delta \theta_{o,\text{obs}}}{2}\right\}
\end{equation} 
where $\delta \theta_{o}$ is the reflection angular sampling. Given \eqref{eq:theta_o_center} and \eqref{eq:theta_o_max}, $\delta \theta_{o}$ is designed to cover the ROI with contiguous and partially overlapped reflection beams. This ensures that each target is illuminated by at least one (possibly more than one) SP-EMS. Herein, we assume that the periodicity $\Lambda$ is exactly double of the illuminated portion of metasurface, i.e., $\Lambda = 2 A_\text{eff}$. This allows exploring all the reflection angles within a single Tx sweep over codebook $\Theta_i$. 
The reflection pattern of the SP-EMS illuminated by the $\ell$th Tx beam is:
\begin{equation}
\begin{split}
R_\ell(\theta_o) \simeq \frac{\sin\left(\pi \frac{d}{\lambda_0} N_\text{mod} \left[ \sin \theta_o - \sin \theta_{o,\ell}\right]\right)}{ 
 \sin\left(\pi \frac{d}{\lambda_0} \left[ \sin \theta_o - \sin \theta_{o,\ell}\right]\right)}
\end{split}
\end{equation}
centered around the reflection angle $\theta_{o,\ell} = \theta_{i,\ell} + \Delta \theta_\text{mod}(x = v \ell \Delta \tau + D \tan \overline{\theta}_{i})$. 
The sampling condition is set by ensuring that contiguous beams are partially overlapped, such that all the ROI is fully covered, yielding:
\begin{equation}\label{eq:deltathetao}
   \delta \theta_o \leq  \frac{1}{2} \min_\ell \left\{\frac{\lambda_0}{N_\text{mod} d \cos\theta_{o,\ell}} \right\} \bigg\lvert_{\theta_{i,\ell} = \overline{\theta}_i} 
\end{equation}
where we evaluate the beamwidth by considering the center of the angular sampling interval $\overline{\theta}_i$. Eq. \eqref{eq:deltathetao} implies a constraint on the minimum number of reflection angles, $|\Theta_o|_{\min} =2 \Delta \theta_{o, \text{obs}}/\delta \theta_o$, that maps into a maximum allowed size of each SP-EMS $N^{\max}_\text{mod}$.
For $ \Lambda = 2A_\text{eff}$, the optimal design of the module size shall satisfy $N_\text{mod} < N^{\max}_\text{mod}$. In this latter case, multiple modules significantly illuminate the target. The resolution increases, and it tends to be higher (or much higher) than the one provided by the single module with effective aperture $A_\text{mod} = N_\text{mod} d$, almost achieving the maximum aperture in \eqref{eq:Aeff}, $A_\text{mod} \simeq N_\text{mod} d$. Differently, choosing comparably large modules, $N_\text{mod} > N^{\max}_\text{mod}$, implies that: \textit{(i)} the ROI is not fully covered, thus there could be targets not directly illuminated by any of the modules, \textit{(ii)} a given target in the ROI is maximally illuminated by reflection off \textit{at most} a \textit{single} module, therefore the image resolution is approximately dictated by $A_\text{mod} \ll A_\text{eff}$, and \textit{(iii)} the position of the target in the final image will undergo an angular shift by $\theta_{i,\ell}-\overline{\theta}_i$. To gain insight, Fig. \ref{fig:module_size} shows an image example of a point target in the ROI for $N_\text{mod} > N^{\max}_\text{mod}$ (Fig. \ref{subfig:large_modules})
$N_\text{mod} < N^{\max}_\text{mod}$ (Fig. \ref{subfig:small_modules}), for $f_0=77$ GHz, $D=5$ m, $\overline{\theta}_i=40$ deg, $\Delta \theta_{i,\text{obs}} = 5$ deg. It is evident that designing few large modules implies a coarse quantization of $\Delta \theta(x)$, leading to a low-quality image. Remarkably, for small modules, the image angular resolution is comparable with the upper bound in which the SP-EMS is purposely configured as a lens to focus in the target location (represented by red dashed lines).

\subsection{Selection of the Spatial Period $\Lambda$}\label{subsec:periodicity}

\begin{figure}[!b]
    \centering
    \subfloat[][$\Lambda = 2$ m]{\includegraphics[width=0.5\columnwidth]{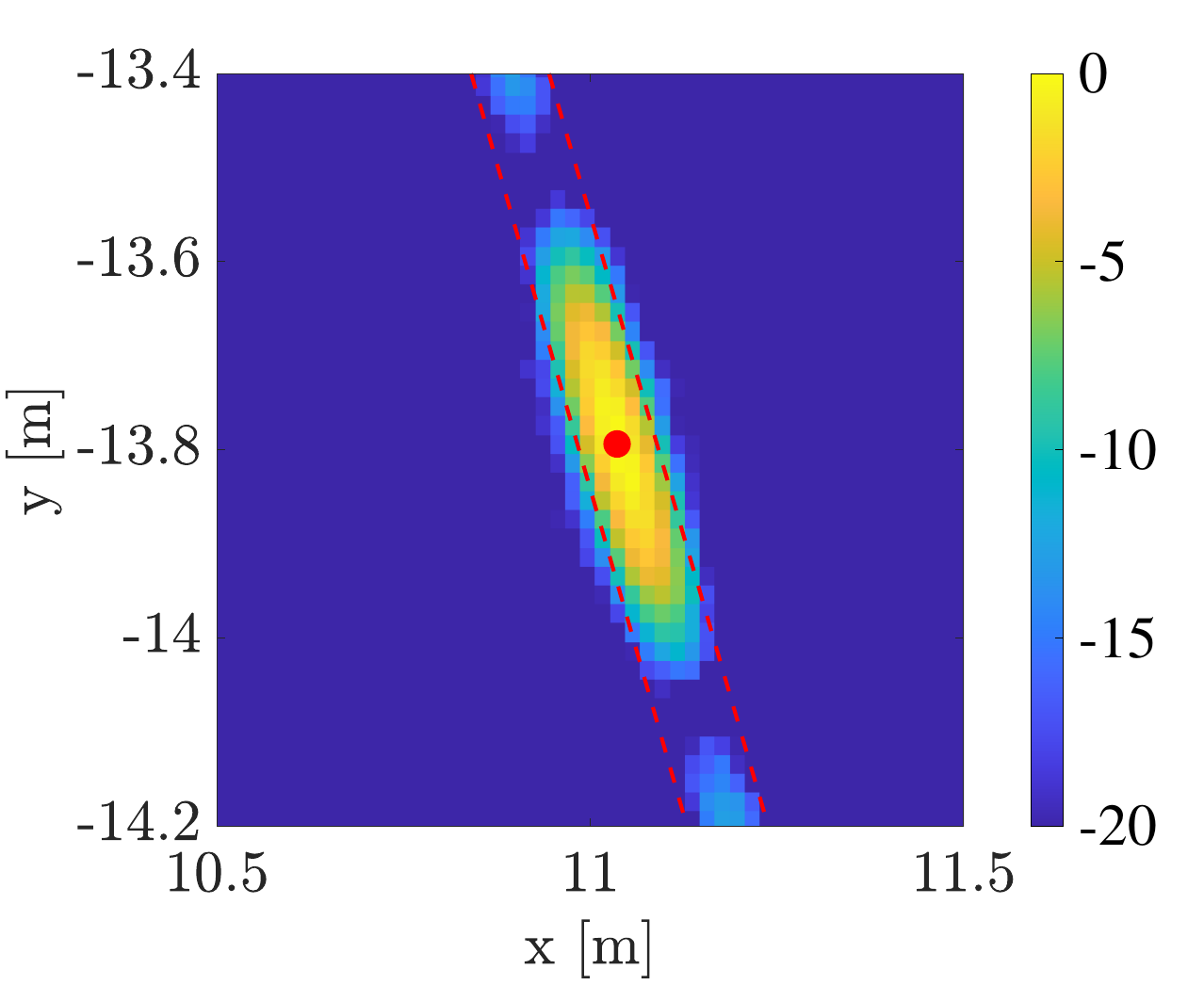}\label{subfig:Lambda1}}
    \subfloat[][$\Lambda = 0.5$ m]{\includegraphics[width=0.5\columnwidth]{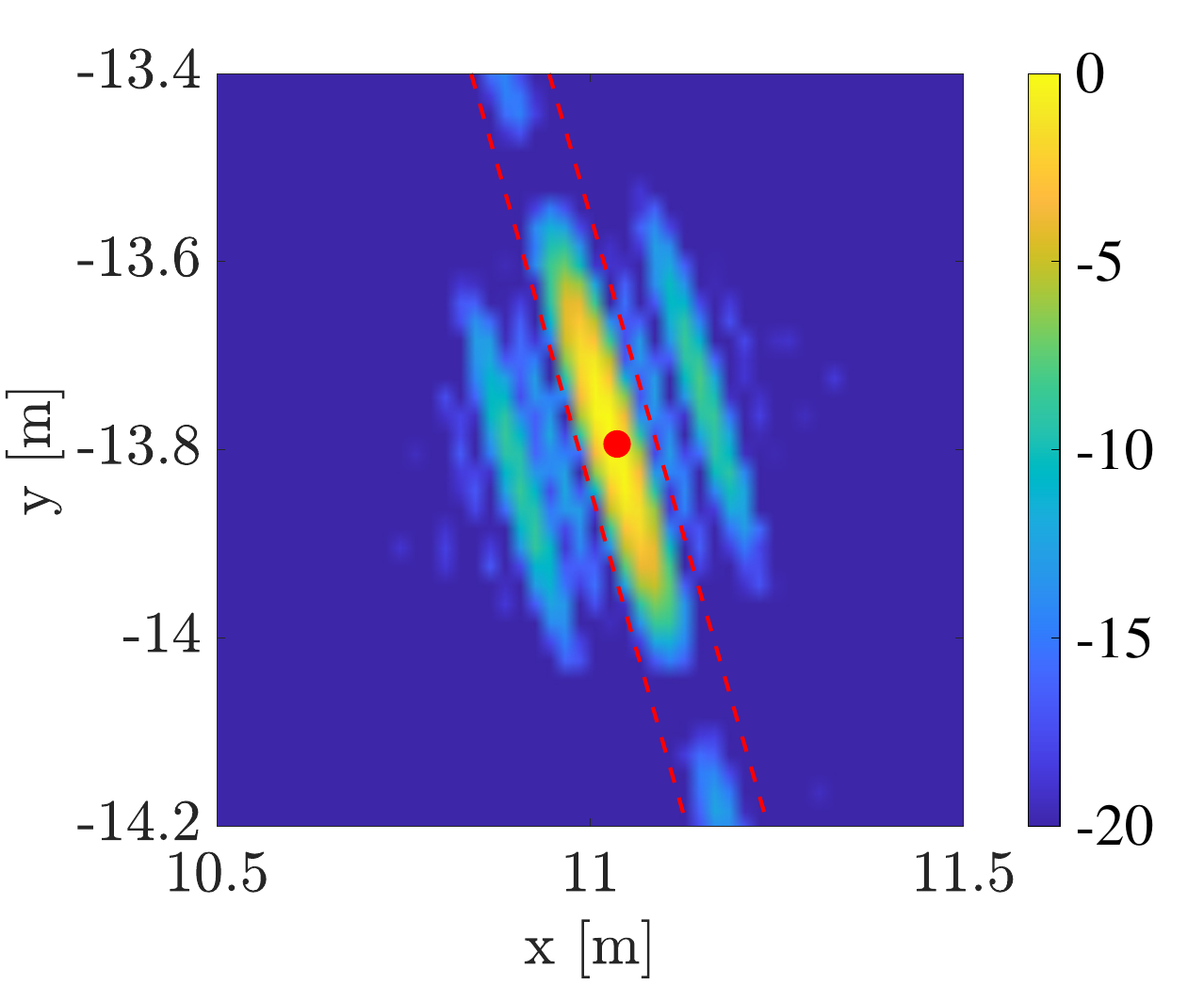}\label{subfig:Lambda05}}
    \caption{Image example of a point target illuminated via a modular SP-EMS whose periodic reflection function has a spatial period (a) $\Lambda = 2$ m and (b) $\Lambda = 0.5$ m. Reducing the periodicity exacerbates the sparse array effect, increasing the sidelobes of the image. }
    \label{fig:periodicity}
\end{figure}

The choice of the spatial period $\Lambda$ involves a trade-off between image resolution (regarded as the width of the main lobe of $H(\mathbf{x})$) and the sidelobes of $H(\mathbf{x})$. For a fixed $\Theta_i$, designing $\Lambda = 2 A_\text{eff}$ ensures an image resolution dictated by an equivalent aperture on the SP-EMS that is comparable with $A_\text{eff}$ (slightly less), with an  acceptable level of sidelobes, as shown in Fig. \ref{subfig:small_modules}. Differently, by increasing the periodicity by a factor $P$, e.g, $\Lambda = 2 A_\text{eff}/P$, the maximum module size reduces accordingly to $N^{\max}_\text{mod} = \Lambda/(P d |\Theta_o|_{\min})$, but the targets are, on average, illuminated by $P$ disjoint apertures within $A_\text{eff}$. Recalling Appendix \ref{sec:resolution}, the image resolution increases, but the sidelobe level rises accordingly as function of the spacing between the two considered modules. 
Moreover, the aforementioned spacing between the reflecting modules changes with the specific target. As a matter of example, we report in Fig. \ref{fig:periodicity} the comparison between the case of $\Lambda = 2$ m and $\Lambda=  0.5$ m, for the very same illumination, $A_\text{eff}=1$ m. A larger value of the periodicity $\Lambda$ leads to an image with good resolution (mostly dictated by $A_\text{eff}$) and low sidelobes ($>20$ dB below the peak), while increasing the periodicity slightly narrows the main lobe but also raises the sidelobes at less than $10$ dB from the peak. Therefore, given $\Theta_i$, the optimum periodicity $\Lambda$ should be designed such that $\Lambda = 2 A_\text{eff}$. 

\section{Effect of Non-Ideality on Sensing}\label{sec:nonidealities}


The image formation by BP described in \eqref{eq:BP} requires that the source has the perfect knowledge of its position in space w.r.t. both the position of the reflection plane as well as the location of the ROI, such that to evaluate the correct distances $D_{i,\ell}$ and $D_{o,\ell}(\mathbf{x})$. In practice, an imperfect knowledge of $D_{i,\ell}$ and/or $D_{o,\ell}(\mathbf{x})$ maps into an image degradation, since the BP yields:
\begin{equation}\label{eq:BP_error_1}
        \widehat{I}(\mathbf{x}) \hspace{-0.1cm}= \hspace{-0.1cm}\sum_\ell y_\ell\left(t \hspace{-0.05cm}= \hspace{-0.05cm}\frac{2\left[\widehat{D}_{i,\ell} \hspace{-0.05cm}+ \hspace{-0.05cm}\widehat{D}_{o,\ell}(\mathbf{x})\right]}{c}\right) \hspace{-0.05cm}e^{j \frac{4 \pi}{\lambda_0} \left[\widehat{D}_{i,\ell} + \widehat{D}_{o,\ell}(\mathbf{x})\right]}.
\end{equation}
In the following, we analyze the effect of two sources of uncertainty on $D_{i,\ell}$ and $D_{o,\ell}(\mathbf{x})$, that one due to an imperfect knowledge of the source-plane distance $D$, and an error on the plane orientation w.r.t. source trajectory. 
Further, we also consider the case in which the source is uncertain on the portion of illuminated metasurface. 

\begin{figure}[!t]
    \centering
    \subfloat[][Sketch of an error  $\varepsilon$ on $D$ ]{\includegraphics[width=0.8\columnwidth]{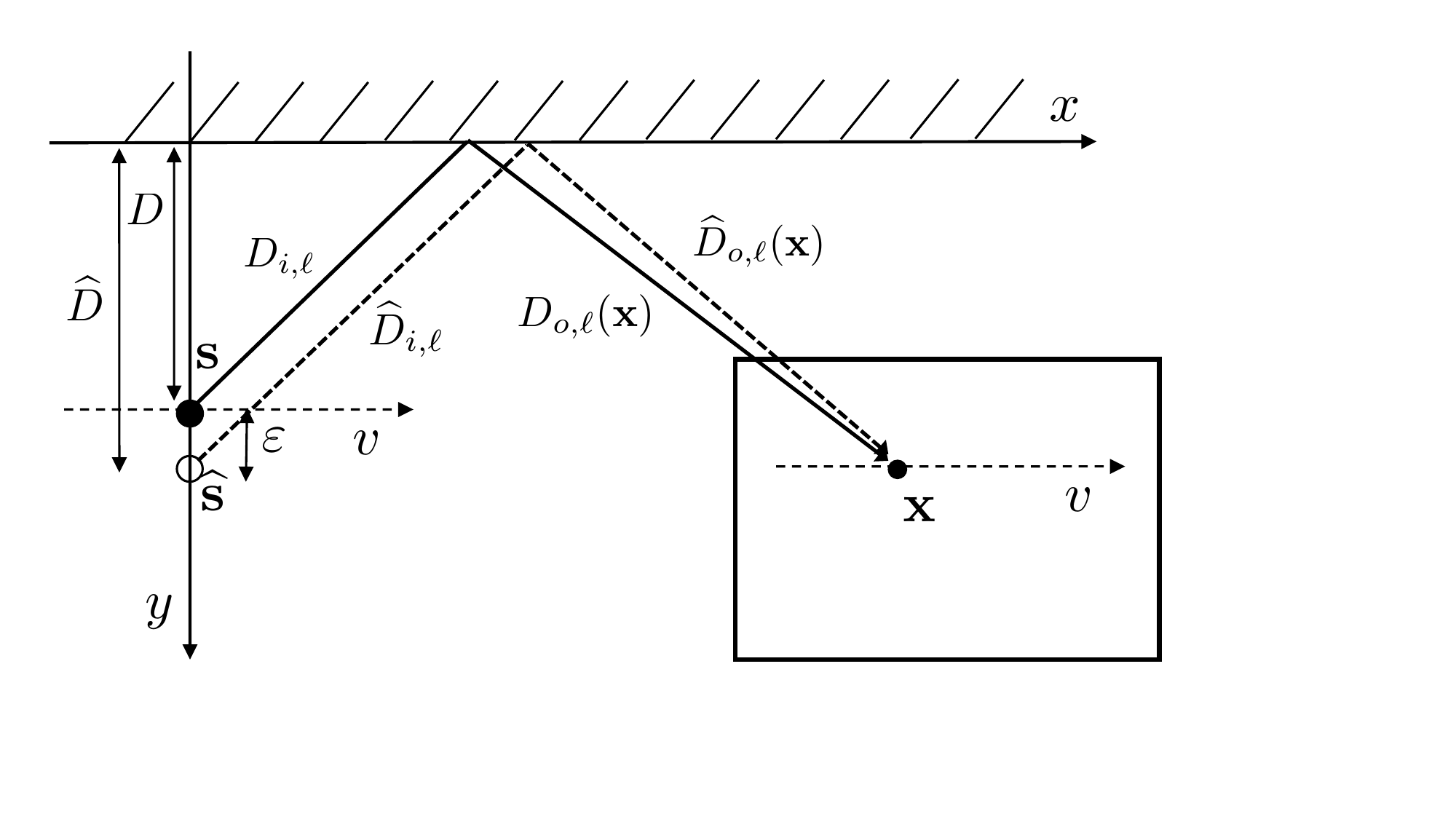}\label{subfig:error_D_sketch}}\\ \vspace{-0.25cm}
    \subfloat[][Distance error vs $\varepsilon$, for different $\varepsilon$. ]{\includegraphics[width=0.8\columnwidth]{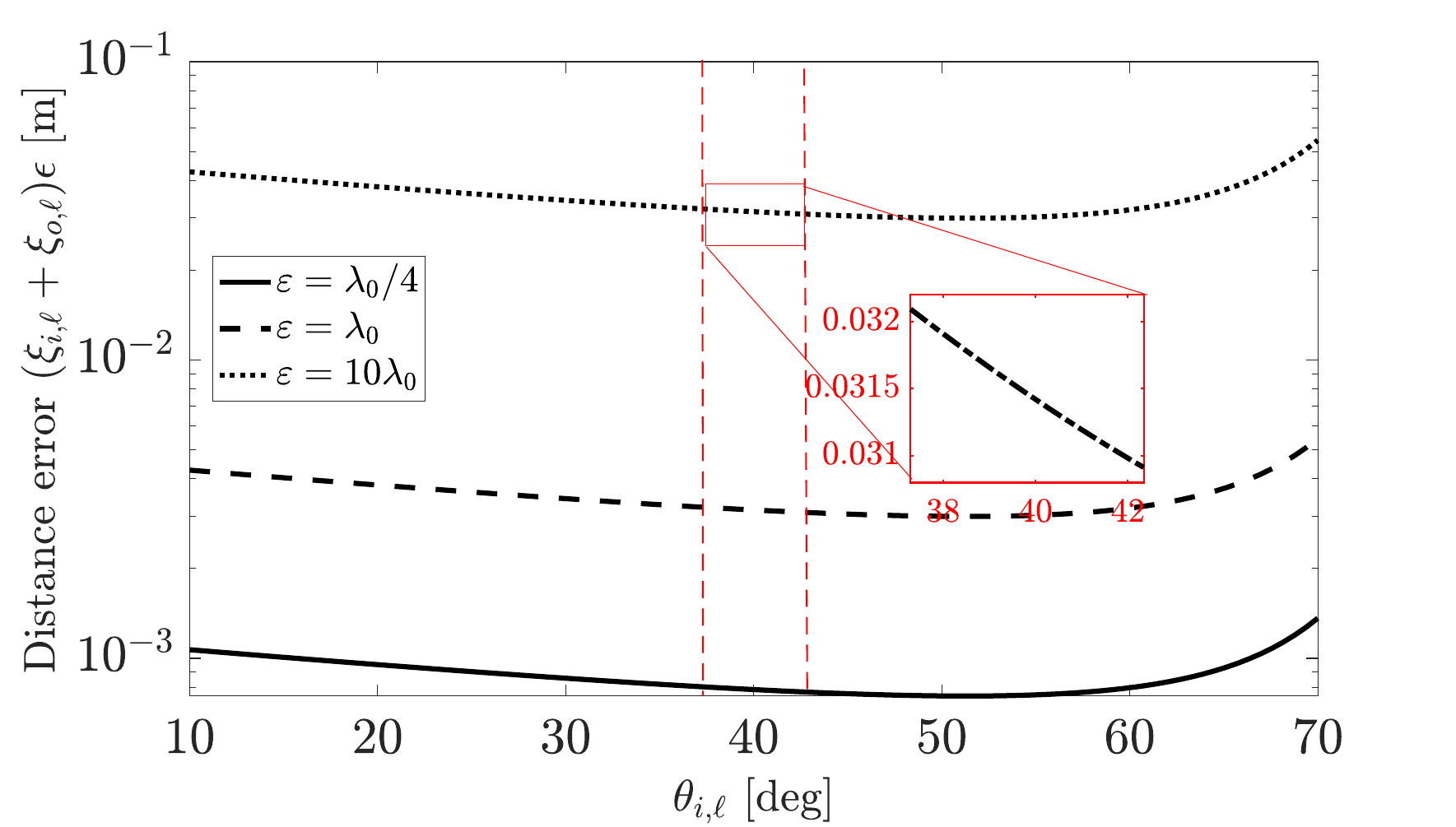}\label{subfig:distance_error_D}}\\ \vspace{-0.25cm}
    \subfloat[][Image example for $\varepsilon = 10 \lambda_0$]{\includegraphics[width=0.5\columnwidth]{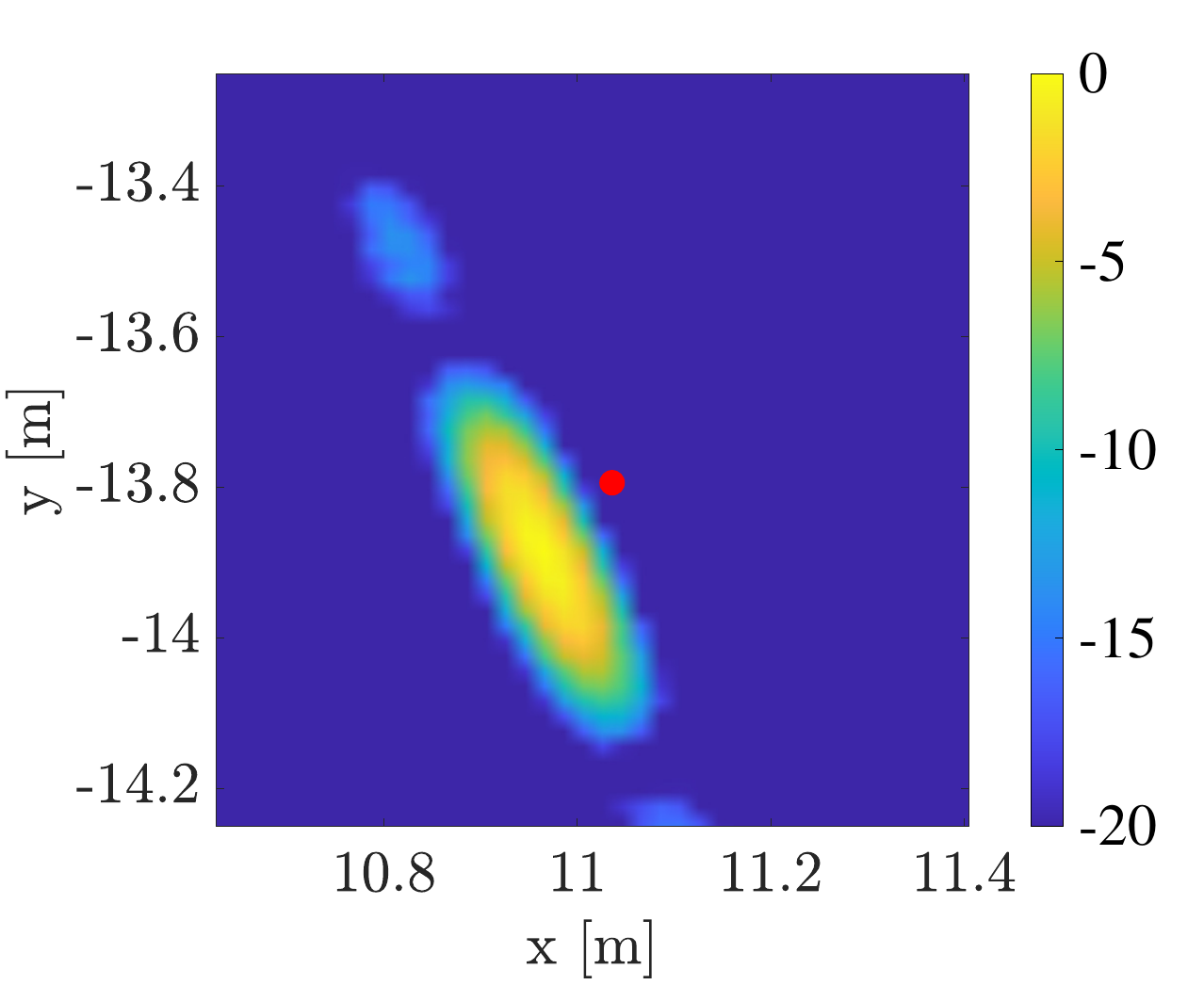}\label{subfig:image_error_D}}
    \caption{Effect of an error $\varepsilon$ on $D$: (a) sketch of the geometry
    (b) overall error in the path-length, varying $\varepsilon$, (c) consequent distorted image of a target obtained for $D = 5$ m, $f_0=77$ GHz, $\overline{\theta}_i = 40$ deg, $\Delta \theta_{i,\text{obs}}=5$ deg. The red inset in (b) zooms over $\Delta \theta_{i,\text{obs}}=5$ deg, showing the linear trend of the distance error. }
    \label{fig:error_D}
\end{figure}

\subsection{Effect of an Error on Distance $D$}\label{subsec:distance}

The first analysis concerns the case in which the source is affected by an error on the distance with the plane $D$, thus $\widehat{D} = D + \varepsilon$. We assume that both source and the ROI move along a linear trajectory parallel to the plane, thus $\widehat{\mathbf{s}}_\ell = (s_x(0)+ \ell v \Delta \tau, D + \varepsilon)$, $\widehat{\mathbf{r}}_\ell = (r_x(0)+ \ell v \Delta \tau, r_y + \varepsilon)$. The estimated propagation distances can be made more tractable by exploring the non-linear relationship: 
\begin{equation}
\begin{split}
 \widehat{D}_{i,\ell} &= D_{i,\ell} + \xi_{i, \ell} \, \varepsilon\\
        \widehat{D}_{o,\ell}(\mathbf{x}) &\simeq D_{o,\ell}(\mathbf{x}) + \xi_{o, \ell}\, \varepsilon
\end{split}
\end{equation}
where $ \xi_{i, \ell} = \sec \theta_{i,\ell}$ and 
\begin{equation}
    \xi_{o, \ell} =  \frac{- \tan \theta_{i,\ell} (x - D \tan \theta_{i,\ell}) + (D - y)}{D_{o,\ell}(\mathbf{x})}.  
\end{equation}
The estimated image according to \eqref{eq:BP_error_1} yields
\begin{equation}\label{eq:image_error_D}
    \widehat{I}(\mathbf{x}) \simeq \sum_\ell \eta_\ell \, g\left(\frac{2 (\xi_{i, \ell} + \xi_{o, \ell})\varepsilon}{c}\right) e^{-j\frac{4 \pi}{\lambda_0} (\xi_{i, \ell} + \xi_{o, \ell})\varepsilon}
\end{equation}
where factor $\eta_\ell$ accounts for path-loss and reflection gain from the plane. An error $\varepsilon$ leads to a equivalent distance error $(\xi_{i, \ell} + \xi_{o, \ell})\varepsilon$ that maps into: \textit{(i)} sampling the base-band signal (after matched filtering) in the wrong time instant, that does not correspond to the maximum, \textit{(ii)} introducing a phase that is, in general, non-linear w.r.t $\ell$. To gain insight on the role of $\varepsilon$, Fig. \ref{fig:error_D} shows the trend of the distance error $(\xi_{i, \ell} + \xi_{o, \ell})\varepsilon$ with $\theta_{i,\ell}$ (varying $\varepsilon$) and an imaging example for $\varepsilon = 10 \lambda_0$, $D = 5$ m, $f_0=77$ GHz, $\mathbf{x} = (20,D)$, $\overline{\theta}_i = 40$ deg, $\Delta \theta_{i,\text{obs}}=5$ deg. A constant error $\varepsilon$ on $D$ maps into almost constant $(\xi_{i, \ell} + \xi_{o, \ell})\varepsilon$ over the Tx angular observation interval $\Delta \theta_{i,\text{obs}}=5$ deg, that provides a range shift in the final image of the same amount (Fig. \ref{subfig:image_error_D}). If $(\xi_{i, \ell} + \xi_{o, \ell})\varepsilon \leq c/B$, i.e., the distance error is below the range resolution of the system, the baseband waveform $g(t)$ is sampled around the peak. However, the variation of $(\xi_{i, \ell} + \xi_{o, \ell})\varepsilon$ within $\Delta \theta_{i,\text{obs}}=5$ deg is sufficient to affect the carrier phase term within the BP, since it is comparable with $\lambda_0$ (red inset of Fig. \ref{subfig:distance_error_D}). The latter effect maps into an image rotation of few degrees, as shown in Fig. \ref{subfig:image_error_D}.

\subsection{Effect of an Error on Plane Tilting w.r.t. Trajectory}\label{subsec:inclination}
\begin{figure}[!t]
    \centering
    \subfloat[][Effect of an error $\beta$ on the plane tilting]{\includegraphics[width=0.8\columnwidth]{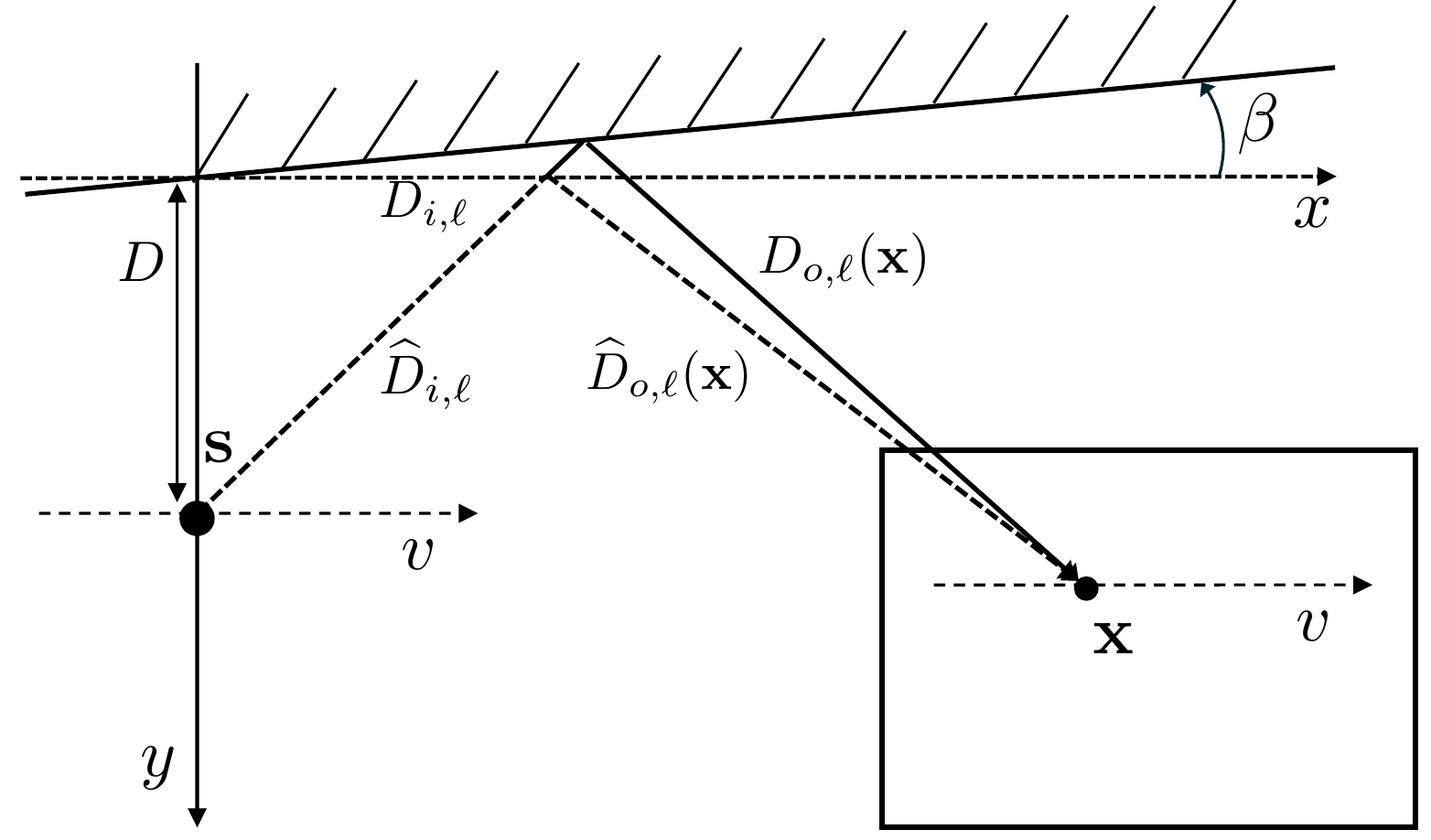}\label{subfig:error_B_sketch}}\\ \vspace{-0.25cm}
    \subfloat[][Distance error for $vT_\text{obs} \simeq 0$, for different $\beta$. ]{\includegraphics[width=0.8\columnwidth]{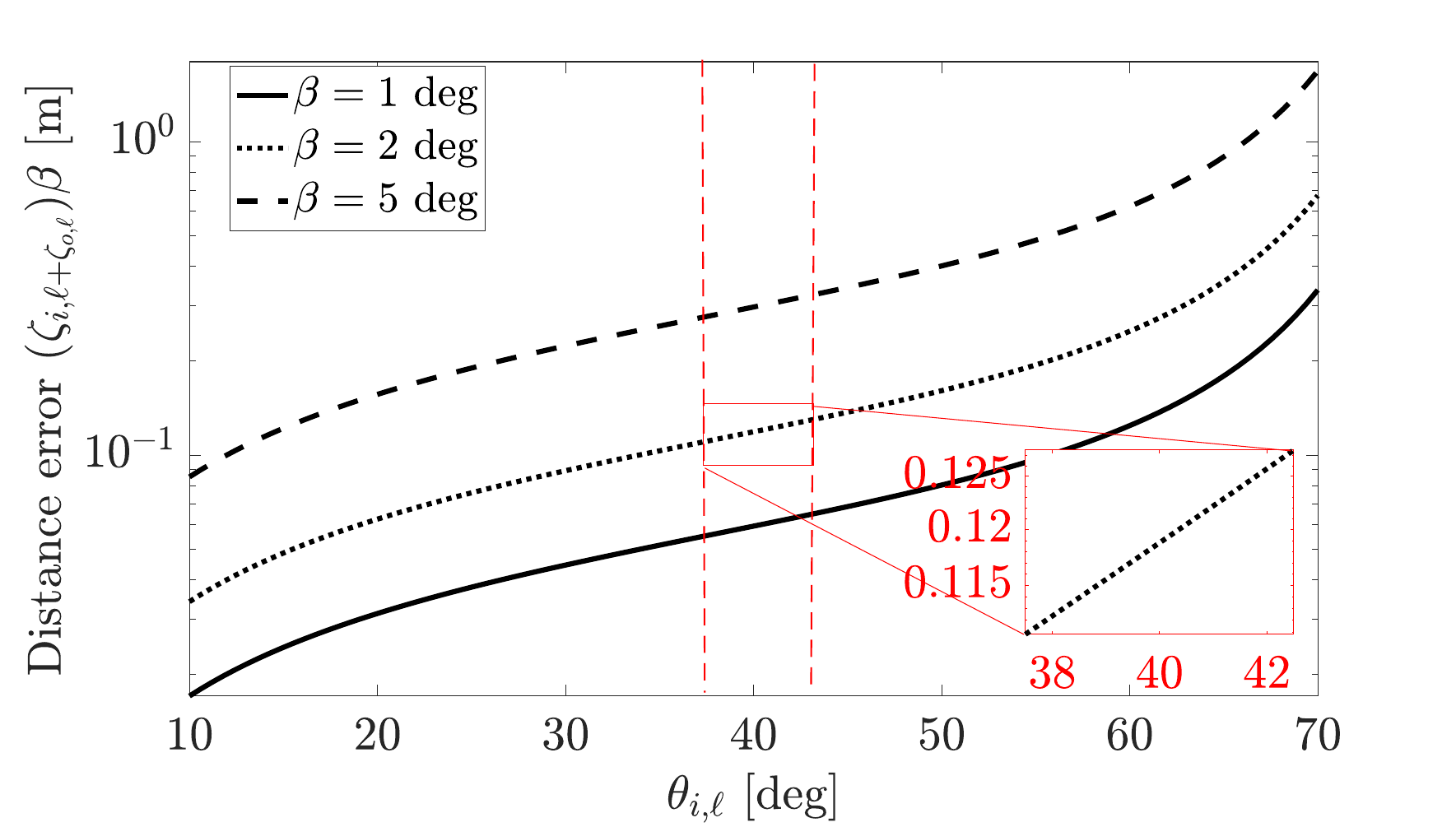}\label{subfig:distance_error_B_instantaneous}}\\ \vspace{-0.25cm}
    \subfloat[][Distance error for $\beta=2$ deg and $vT_\text{obs} > 0$. ]{\includegraphics[width=0.8\columnwidth]{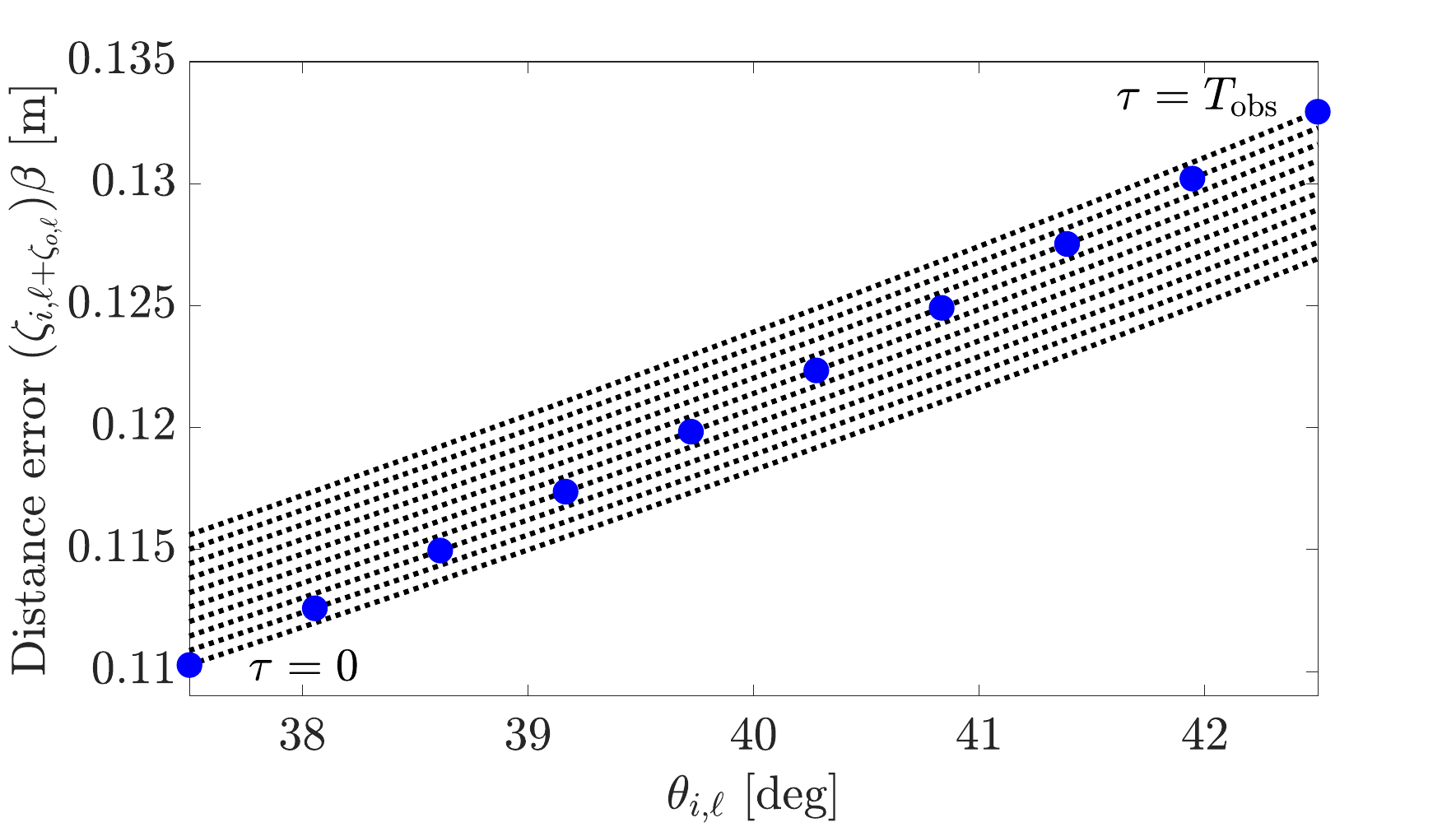}\label{subfig:distance_error_B}}\\ \vspace{-0.25cm}
    \subfloat[][Image example for $\beta=2$ deg]{\includegraphics[width=0.5\columnwidth]{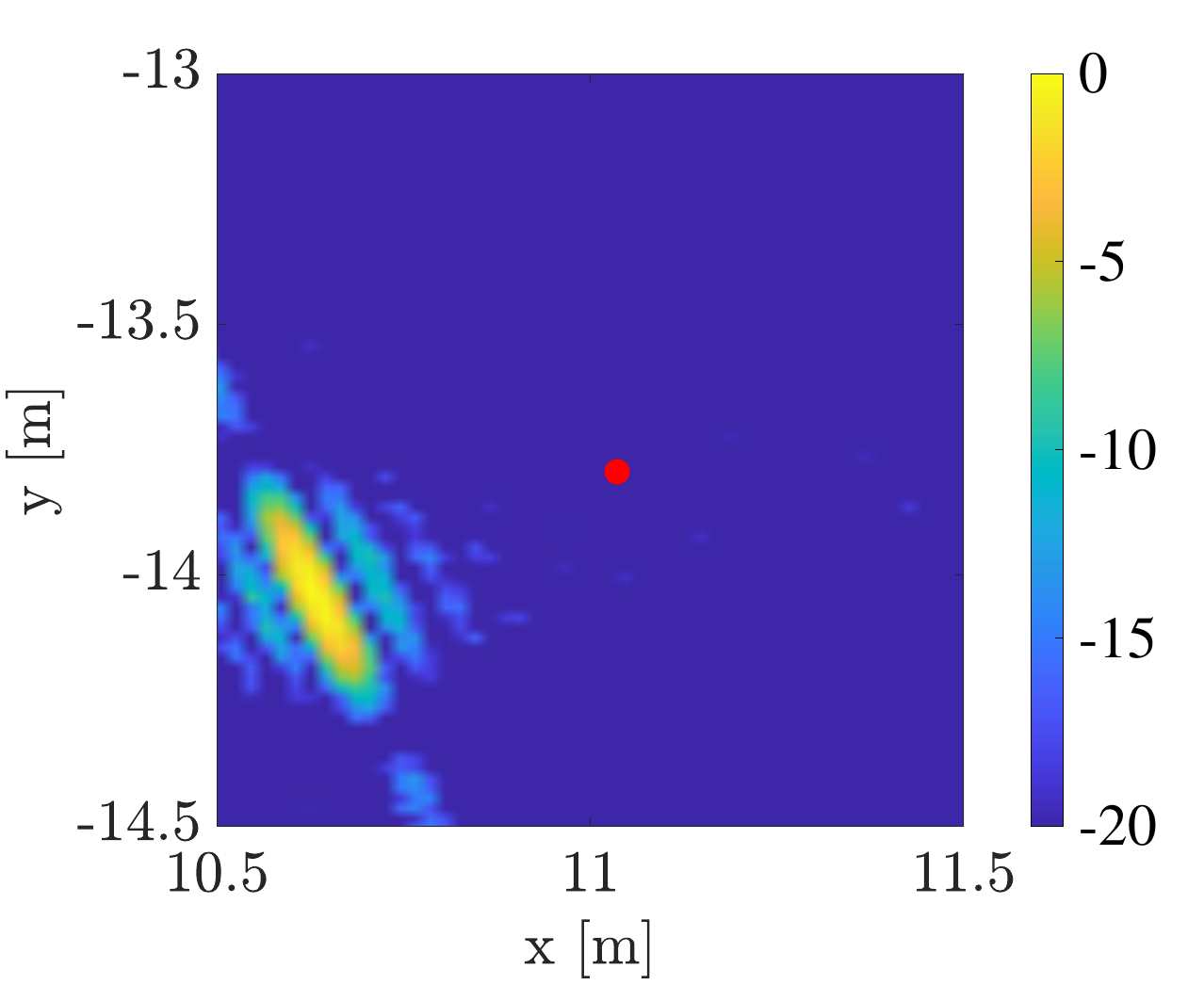}\label{subfig:image_error_B_2}}
    \caption{Effect of an error on the plane inclination $\beta$: (a) sketch of the non-ideal acquisition geometry (b) distance error for a nearly instantaneous sweeping ($v T_\text{obs} \simeq 0$) (c) distance error for $vT_\text{obs} > 0$, over a $\Delta \theta_{i,\text{obs}}=5$ deg, assuming $\beta=2$ deg and $|\Theta_i|=10$ (blue dots represent the effective distance error), (d) image example for $\beta=2$ deg.}
    \label{fig:error_B}
\end{figure}

The second analysis concerns the case in which the source is travelling diverging from the reflection plane, thus there is an error on the orientation of the plane w.r.t. its trajectory. An angular error $\beta$ on the tilting of the reflection plane implies an equivalent tilted motion of the source and the ROI, $\widehat{\mathbf{s}}_\ell = (s_x(0)+ \ell v \Delta \tau \cos \beta, D - \ell v \Delta \tau \sin \beta)$ and $\widehat{\mathbf{r}}_\ell = (r_x(0)+ \ell v \Delta \tau \cos \beta, r_y(0) - \ell v \Delta \tau \sin \beta)$. Now, the distances are approximated at the first order as:
\begin{equation}\label{eq:BP_error_2}
\begin{split}
 \widehat{D}_{i,\ell} &\simeq D_{i,\ell} + \zeta_{i, \ell} \, \beta\\
        \widehat{D}_{o,\ell}(\mathbf{x}) &\simeq D_{o,\ell}(\mathbf{x}) + \zeta_{o, \ell}\, \beta
\end{split}
\end{equation}
where 
\begin{equation}\label{eq:zeta_i}
    \zeta_{i, \ell} = \left(D\tan\theta_{i,\ell} + \ell v\Delta \tau\right)\sqrt{1+\tan^2\theta_{i,\ell}}
\end{equation}
and $\zeta_{o, \ell}$ is reported in \eqref{eq:zeta_o}.
\begin{figure*}
\begin{equation}\label{eq:zeta_o}
\begin{split}
        \zeta_{o, \ell} = D_{o,\ell}(\mathbf{x})\tan\theta_{i,\ell}- \frac{\left(x-D\tan\theta_{i,\ell}\right)\left(x+\ell v \Delta\tau\right)\tan\theta_{i,\ell} + \left(D-y\right)\left(y\tan\theta_{i,\ell}+\ell v\Delta\tau\right)}{D_{0,\ell}(\mathbf{x})} 
\end{split}
\end{equation}\hrulefill
\end{figure*}
Similarly to \eqref{eq:image_error_D}, the estimated image is
\begin{equation}\label{eq:image_error_B}
    \widehat{I}(\mathbf{x}) \simeq \sum_\ell \eta_\ell \, g\left(\frac{2 (\zeta_{i, \ell} + \zeta_{o, \ell})\beta}{c}\right) e^{-j\frac{4 \pi}{\lambda_0} (\zeta_{i, \ell} + \zeta_{o, \ell})\beta}
\end{equation}
where the same considerations outlined for \eqref{eq:image_error_D} apply. Both $\zeta_{i, \ell}$ and $\zeta_{o, \ell}$ explicitly depend on the slow time index sample $\ell$ through the speed $v$. Increasing $v$ (or the Tx codebook size $|\Theta_i|$) the mismatch between the expected distances $\widehat{D}_{i,\ell}/\widehat{D}_{o,\ell} $ and the true ones $D_{i,\ell}/D_{o,\ell}$ increases accordingly, as can be devised from Fig. \ref{subfig:error_B_sketch}. The quantitative effect is instead shown in Fig. \ref{subfig:distance_error_B_instantaneous}, obtained for $f_0=77$ GHz, $D=5$ m, $\mathbf{s}=(0,D)$, $\mathbf{x} = (20,D)$, $\overline{\theta}_i = 40$ deg, $\Delta \theta_{o,\text{obs}}=5$ deg and $v T_\text{obs} \simeq 0$ (i.e., a nearly instantaneous Tx sweeping for which $\ell v \Delta \tau \approx 0$ in \eqref{eq:zeta_i}-\eqref{eq:zeta_o}). Similarly to Fig. \ref{subfig:distance_error_D}, the distance error is linear within $\Delta \theta_{o,\text{obs}}$, only affecting the carrier phase. The final image of the target is thus subject to an angular shift, as shown in Fig. \ref{subfig:image_error_B_2}, obtained for $\beta=2$ deg. By increasing the velocity, for which $v T_\text{obs} > 0$, the effective distance error increases accordingly, as shown in Fig. \ref{subfig:distance_error_B}.

\subsection{Uncertainty on the Portion of Illuminated Plane}\label{subsec:uncertainty}

\begin{figure}[!t]
     \centering
     \subfloat[][Sketch of a random illumination of the reflection plane]{\includegraphics[width=0.8\columnwidth]{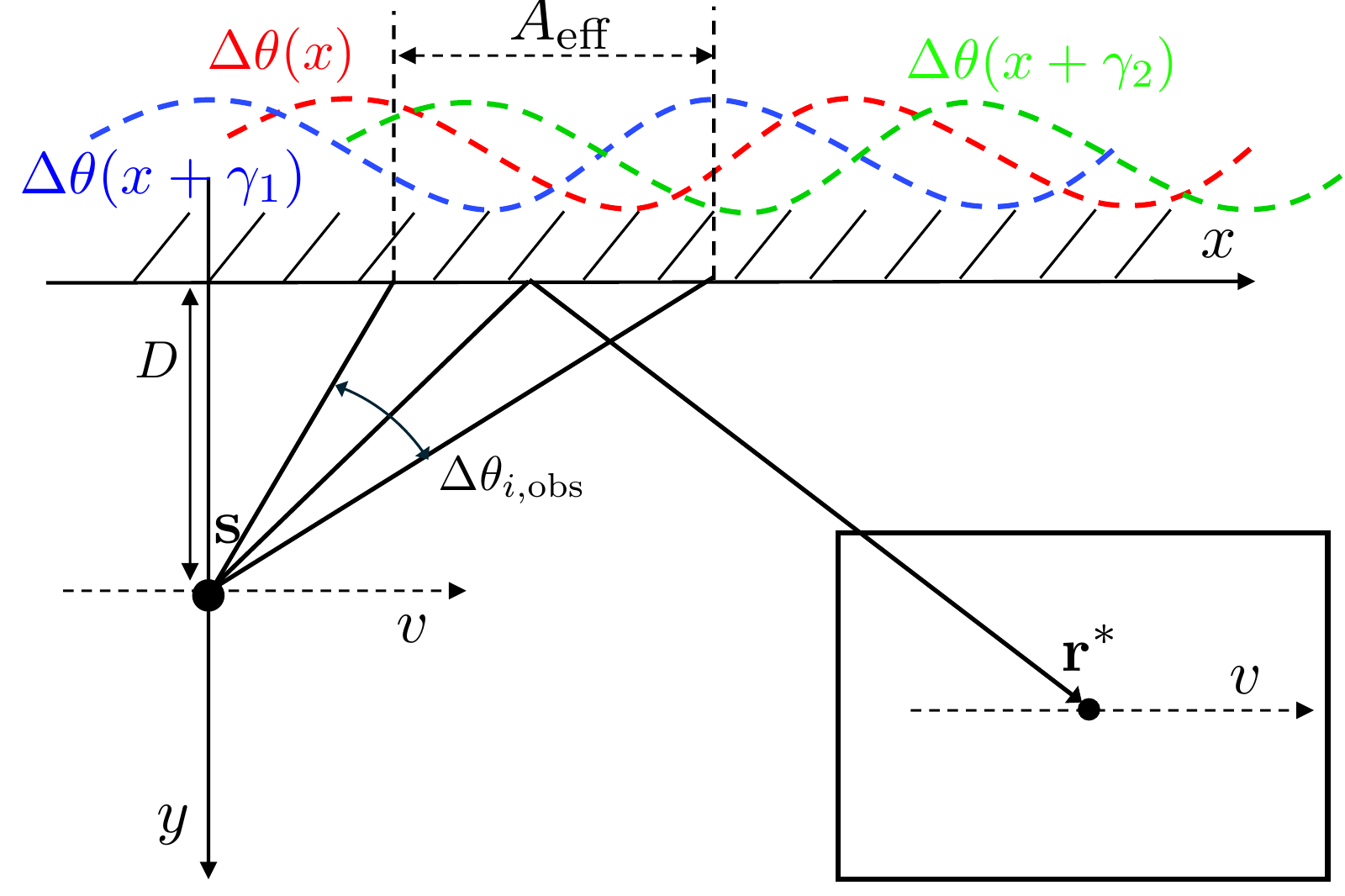}\label{subfig:error_gamma_sketch}}\\
      \subfloat[][]{\includegraphics[width=0.4\columnwidth]{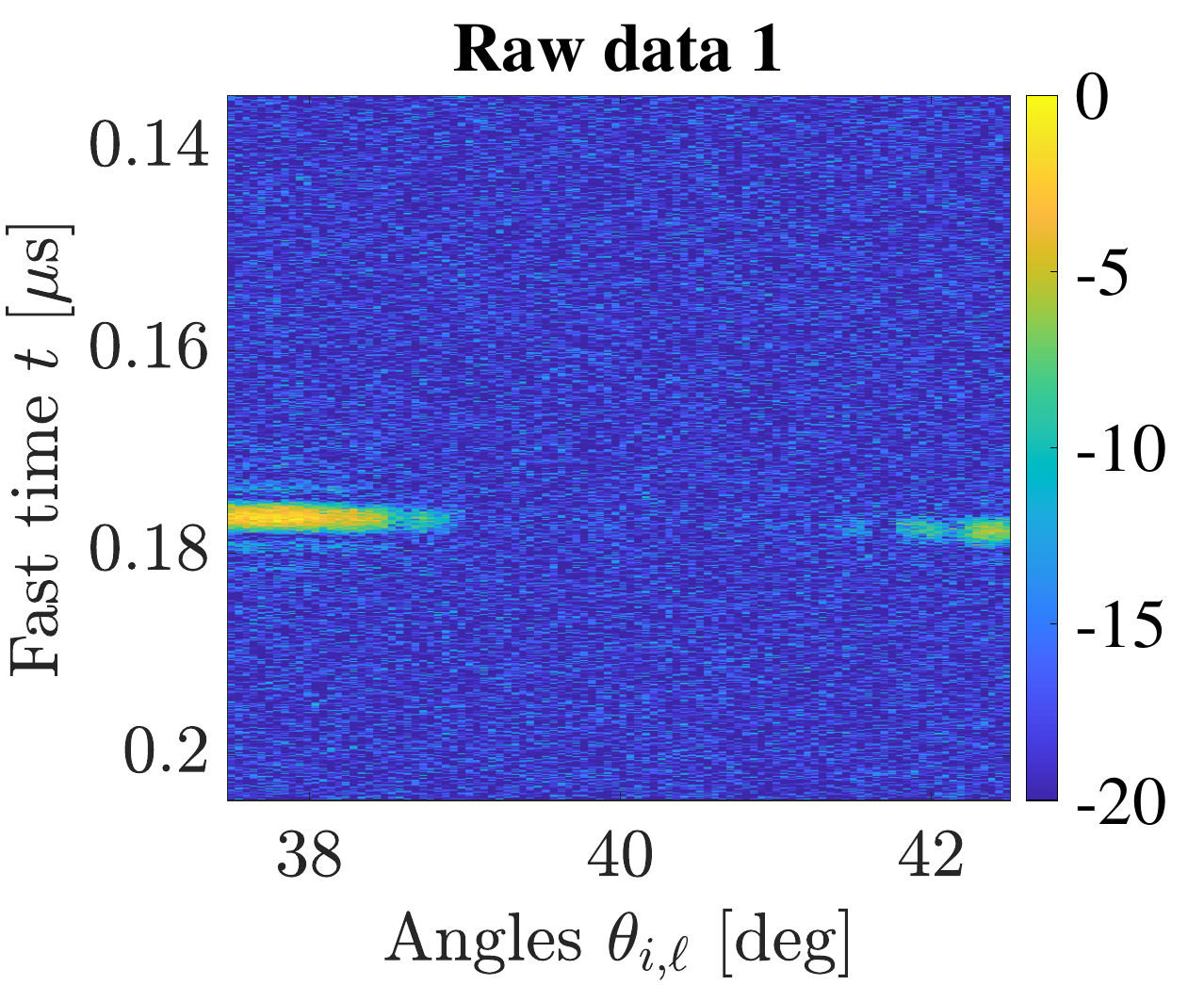}\label{subfig:RC1}}
      \subfloat[][]{\includegraphics[width=0.4\columnwidth]{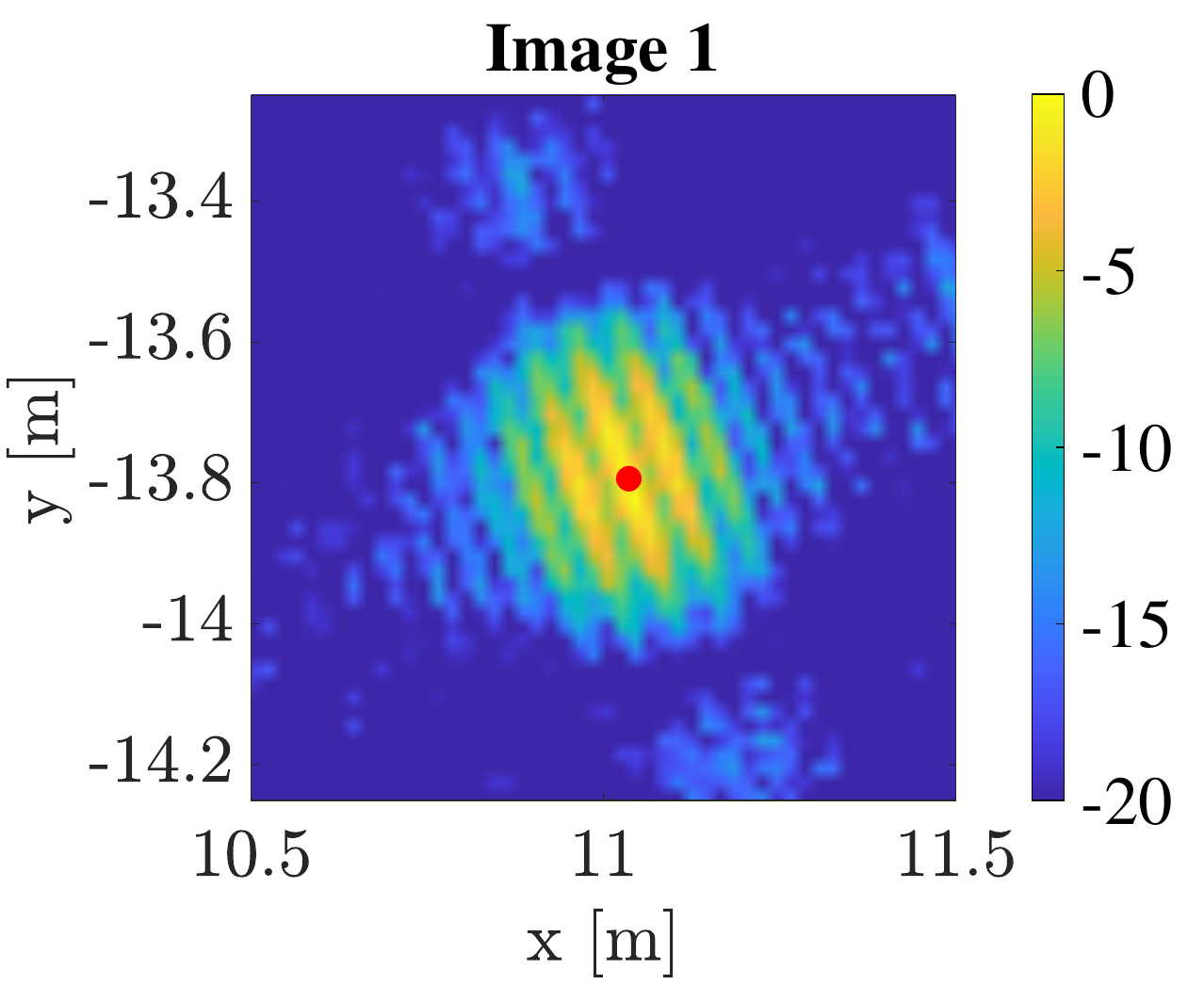}\label{subfig:image1}}\\ \vspace{-0.25cm}
      \subfloat[][]{\includegraphics[width=0.4\columnwidth]{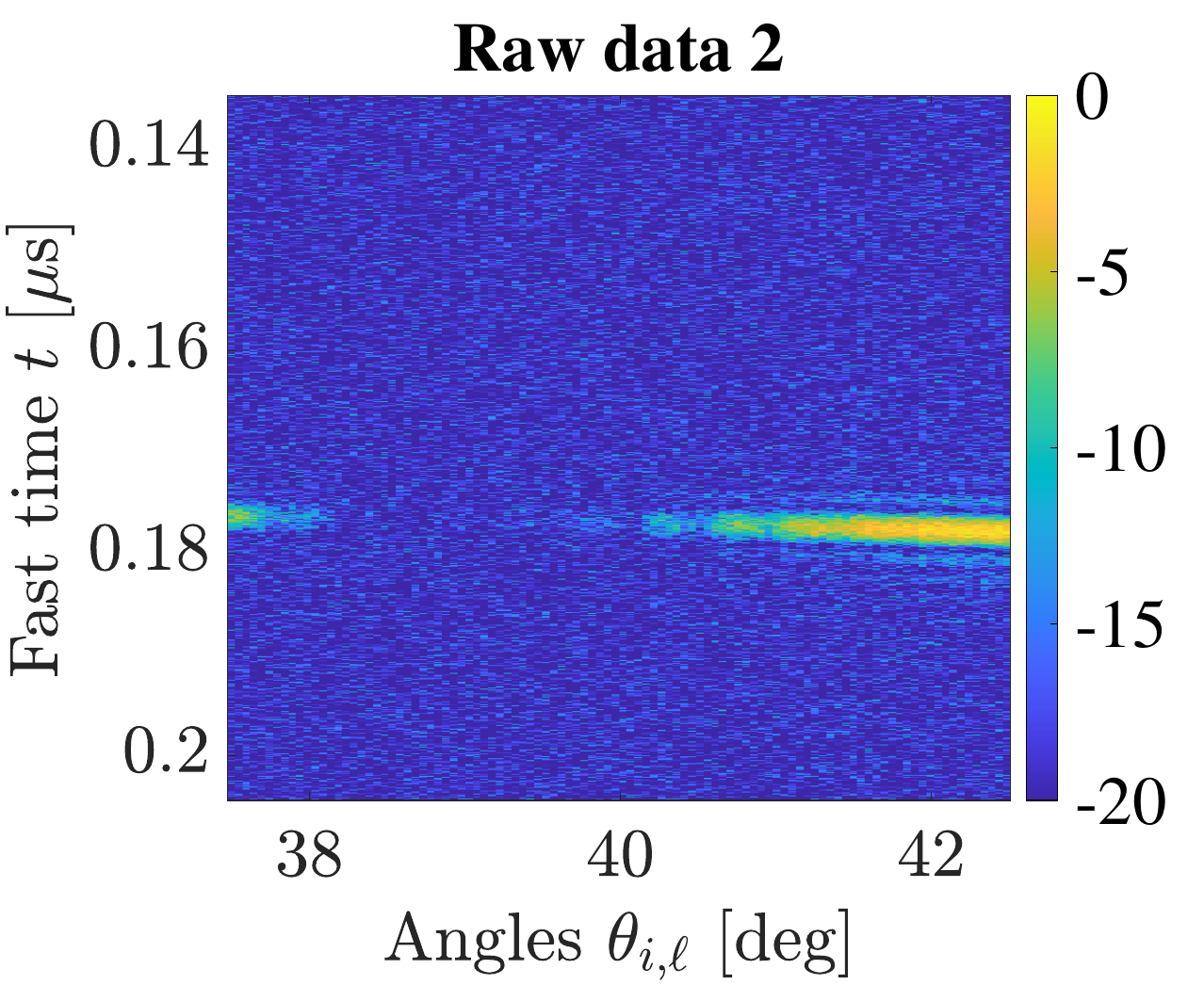}\label{subfig:RC2}}
     \subfloat[][]{\includegraphics[width=0.4\columnwidth]{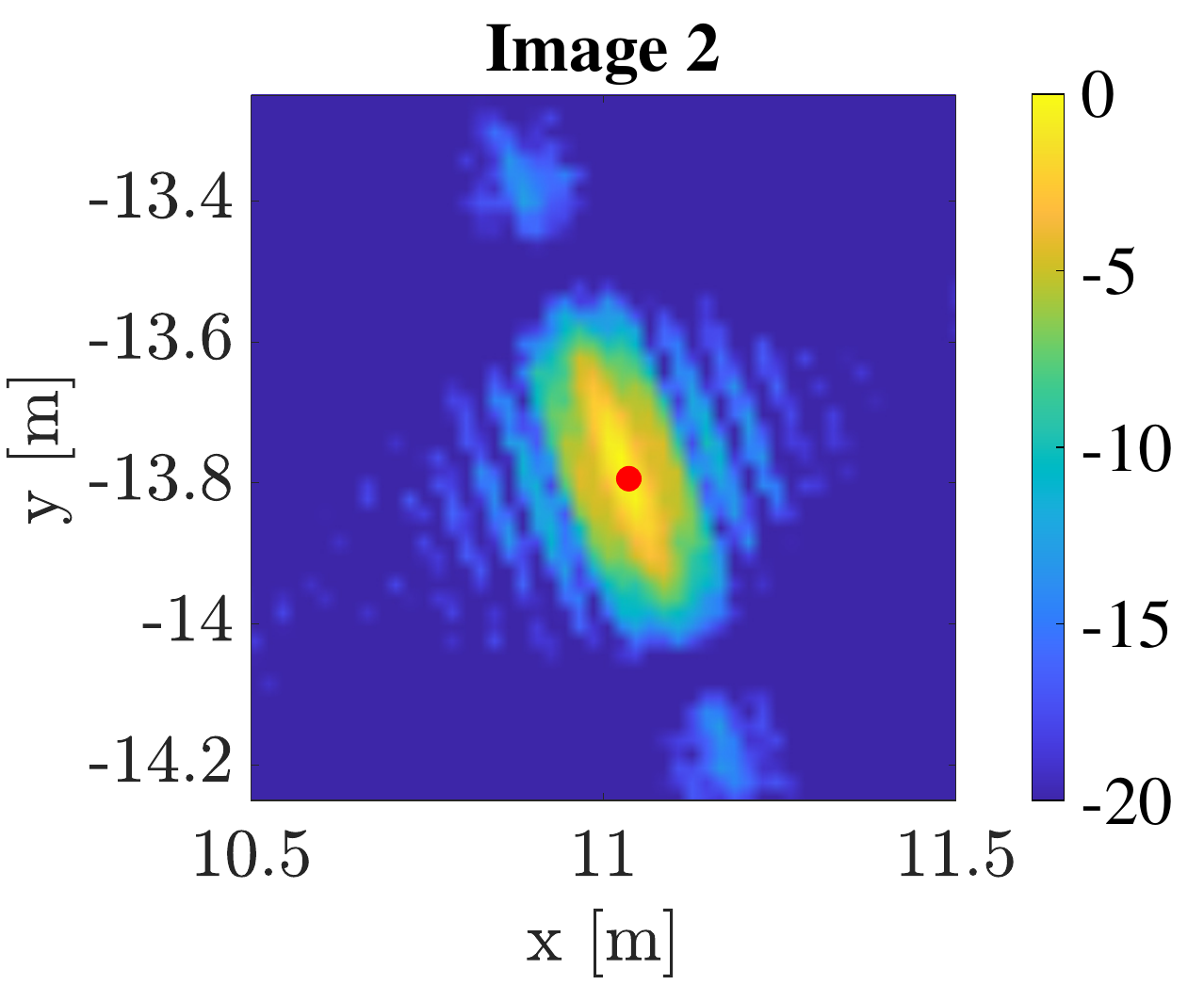}\label{subfig:image2}}\\ \vspace{-0.25cm}
      \subfloat[][]{\includegraphics[width=0.4\columnwidth]{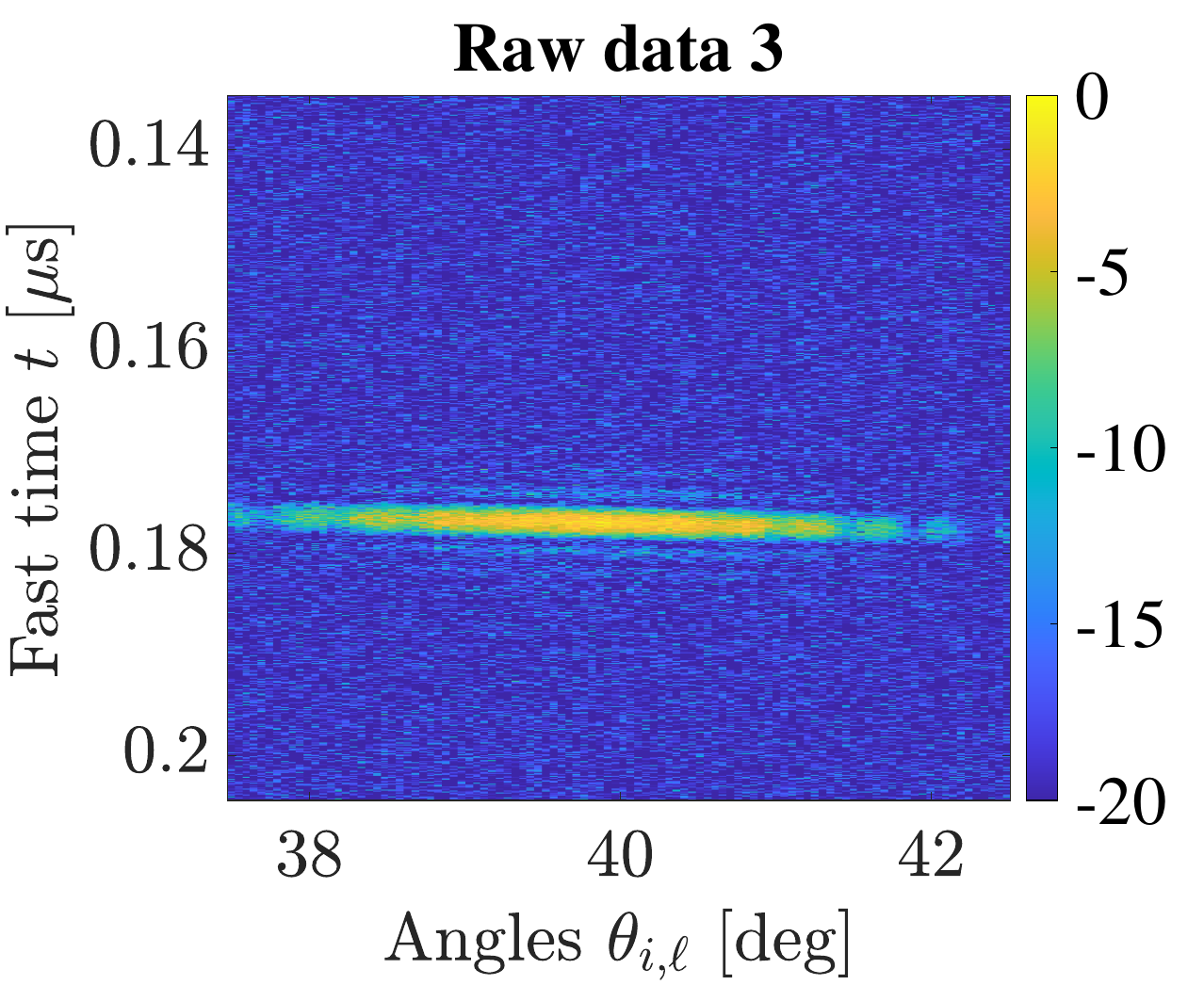}\label{subfig:RC3}}
      \subfloat[][]{\includegraphics[width=0.4\columnwidth]{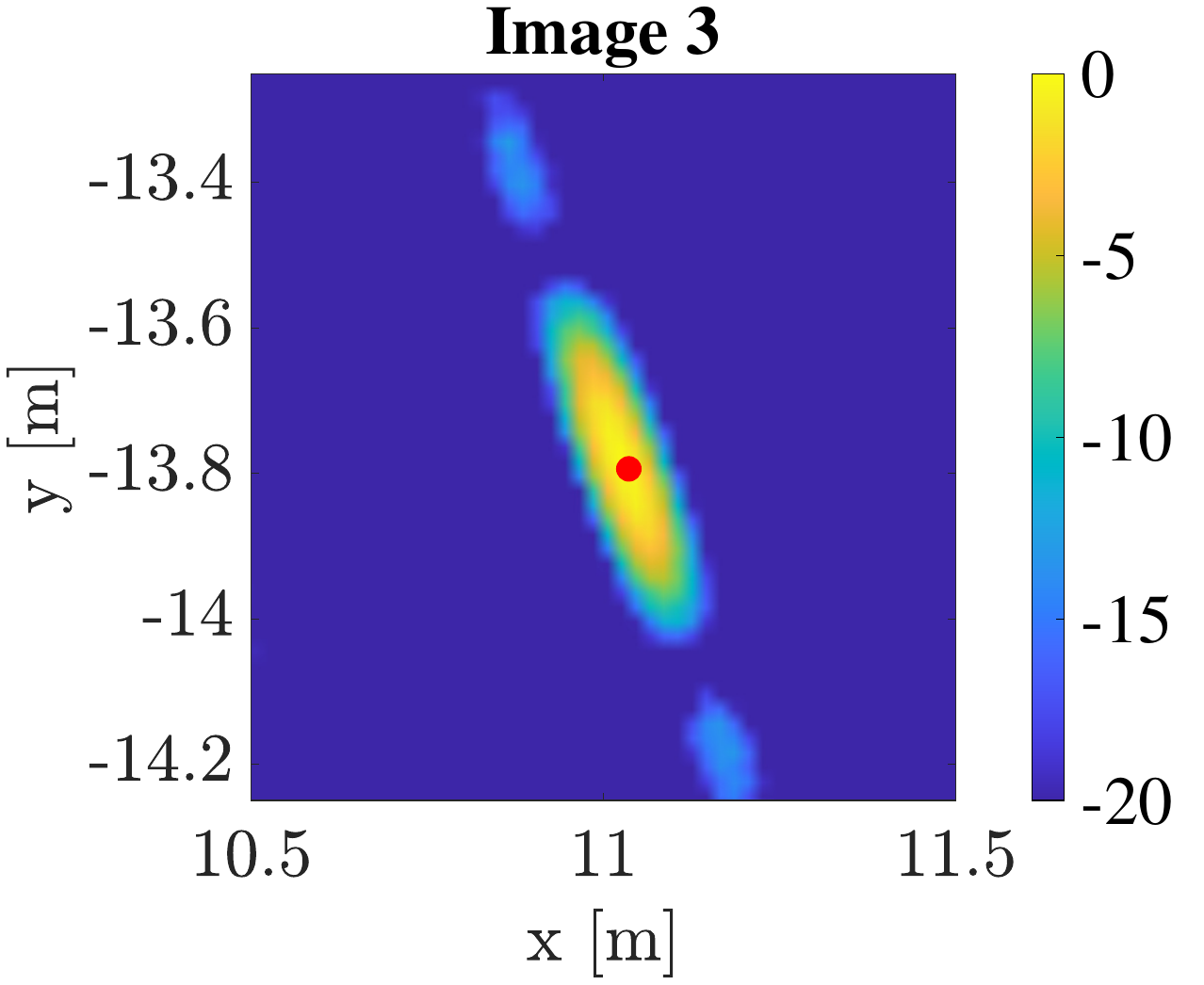}\label{subfig:image3}}
     \caption{Effect of a random illuminated portion of the reflection plane: (a) sketch of the issue (c,e,g) imaging example for 3 values of $\gamma$, i.e., for a random plane illumination by the source, corresponding to (b,d,f) raw data over the fast-time, and slow-time. Different images are obtained as a consequence of the effective portion of the plane that illuminates the target. }
    \label{fig:multisnap}
  \end{figure}

So far, we have considered the case in which the source illuminates a precise portion of the reflection plane $2A_\text{eff} = \Lambda$ (half a cycle of the spatial reflection sinusoid), such that to explore \textit{all} the reflection angles $\theta_o\in\Theta_o$ within a single beam sweeping $\theta_i\in\Theta_i$. For $\Delta\theta(x)$ set as \eqref{eq:angle_periodic_design} (a cosinusoid), the source in $\mathbf{s}=(0,D)$ explores the whole set of reflection angles if it illuminates the plane from $x=0+ 2 \pi a $ to $x=\Lambda/2+2 \pi a$, for any integer $a>0$. In this latter case, the image quality can be accurately predicted, as shown in Section \ref{sec:system_design}. However, the source is typically not \textit{spatially synchronized} with the reflection cosinusoid, thus it illuminates a random portion $2A_\text{eff} = \Lambda$ of the plane, with an effective reflection function  
\begin{equation}\label{eq:angle_periodic_random}
    \widetilde{\Delta\theta}(x) = (\overline{\theta}_o \hspace{-0.05cm}- \hspace{-0.05cm}\overline{\theta}_i) \hspace{-0.05cm}+\hspace{-0.05cm} \frac{\Delta\theta_{o,\text{obs}}}{2}\cos\left(\frac{2 \pi}{\Lambda} x \hspace{-0.05cm}+ \hspace{-0.05cm}\gamma \right)\bigg\rvert_{x=0}^{x = \frac{\Lambda}{2}}
\end{equation}
where $\gamma$ is a random spatial shift modelling the uncertainty on the source position w.r.t. the plane. A sketch of the aforementioned effect it shown in Fig. \ref{subfig:error_gamma_sketch}. The overall effect is to obtain a different image of the same target for different different Tx sweeps, as shown in Fig. \ref{fig:multisnap}. The latter reports three images obtained by sweeping the Tx codebook over a random portion of the spatial reflection sinusoid, according to \eqref{eq:angle_periodic_random}, together with the acquired raw data $y_\ell(t)$, $t\in[0,\Delta \tau]$, $\ell=1,...|\Theta_i|$. Each image is a function of the effective portion of the illuminated SP-EMS that illuminates the target, that varies with the considered example. In particular, we notice the high image sidelobes in the first example (Figs. \ref{subfig:RC1} and \ref{subfig:image1}) as a consequence of the disjoint sets of reflection angles that effectively contribute to the Rx signal. The only opportunity to avoid the random image effect is to coherently combine multiple images, i.e., over consecutive Tx sweeps. 


\section{Results}
\label{sec:results}
In this section, we assess the performance of our proposed stroboscopic sensing system in terms of imaging quality of a complex target shape.  We consider a source operating at $f_0 = 77$ GHz, over a bandwidth of $B=500$ MHz, and emitting the Tx signal over a beamwidth of $\Delta \theta_\text{source}=0.5$ deg. The source is located at distance $D = 5$ m from the reflection plane, with the main beam oriented towards  $\overline{\theta}_{i} = 40$ deg, and sweeping within a codebook of size $\Delta\theta_{i,obs} = 5$ deg. The source moves at constant speed $v = 20$ m/s along $x$, and the PRI is $\Delta \tau= 50$ $\mu$s. The reflection plane is made by a sequence of adjacent modules, where the periodic reflection angle pattern  \eqref{eq:angle_periodic_design_quant} spans a codebook with cardinality $|\Theta_o|=13$ over a spatial periodicity $\Lambda = 2$ m, leading to an horizontal size $N_\text{mod}d = 8$ cm for each module, pre-configured as detailed in Section \ref{sec:system_design}. The ROI is located in $\mathbf{r}^*= (13.8, 11)$ m, of $\Delta_x = \Delta_y = 1$m.
The useful signal at the Rx side is corrupted by thermal noise with power $\sigma_w^2=-87$ dBm. 

\begin{figure}
     \centering
     \subfloat[][Proposed stroboscopic system.]{\includegraphics[width=0.5\columnwidth]{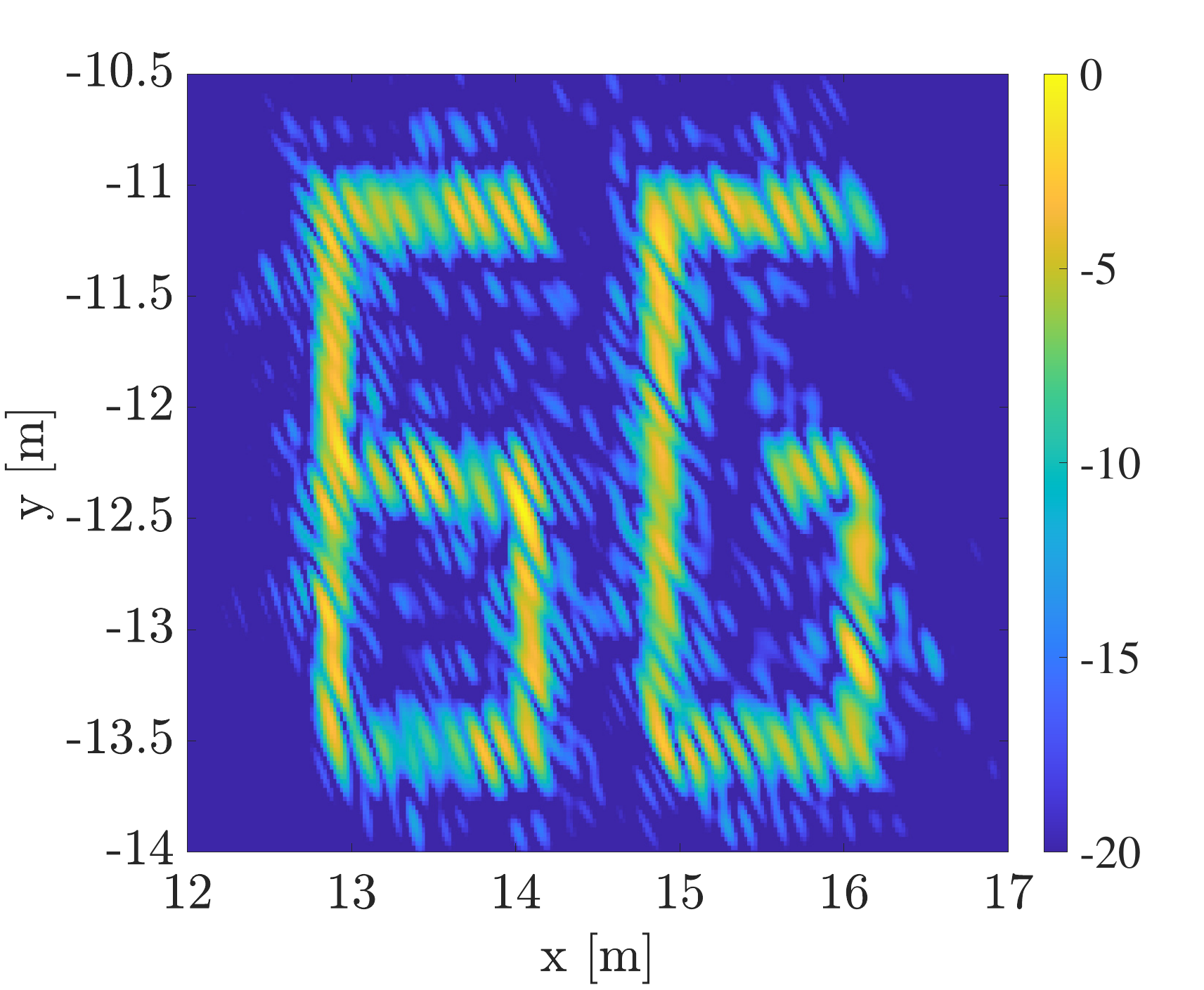}\label{subfig:6G_strobo}} 
     \subfloat[][Mirror \cite{9468353}]{\includegraphics[width=0.5\columnwidth]{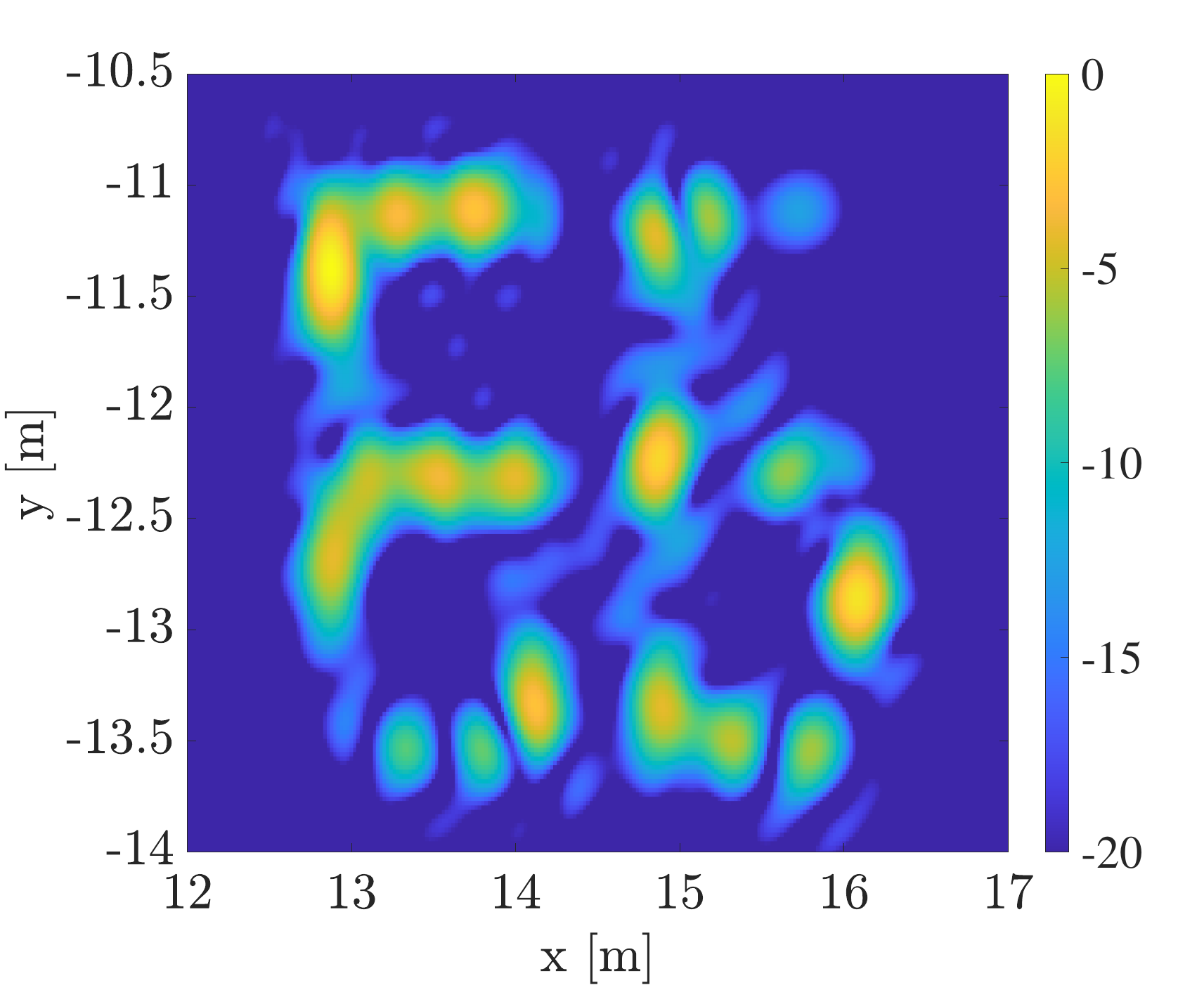}\label{subfig:6G_mirror}}
     \caption{Image example of a complex target shape obtained with (a) the proposed stroboscopic system and (b) a metallic mirror, as considered in \cite{9468353}. Images are normalized to the same maximum value. The proposed system provides remarkable benefits in terms of resolution and coverage over the ROI (all the targets are imaged with the same quality).}
     \label{fig:6G}
\end{figure}

The first result is an exemplary image of a target of complex shape, shown in Fig. \ref{fig:6G}. We compare the proposed stroboscopic sensing system (Fig. \ref{subfig:6G_strobo}) to a mirror-based imaging system, as considered in \cite{9468353} (Fig. \ref{subfig:6G_mirror}), where the source's main beam is pointed such that to maximally illuminate the center of the ROI. The images are both normalized to their maximum values, and the blockage effect is not considered for simplicity. In case of a mirror, some parts of the extended target are not visible in the final image, and the overall image quality is degraded. Moreover, the resolution is dictated by the source's aperture. 
Differently, our system allows sweeping the source's beam over a purposely configured anomalous mirror, such that the resolution is ruled by the illuminated portion across the reflection plane. Now, each target within the ROI is imaged with the same resolution, and the latter is substantially improved compared to the usage of a mirror. The target's shape is clearly distinguishable.   


\begin{figure}
    \centering
    \includegraphics[width=0.8\columnwidth]{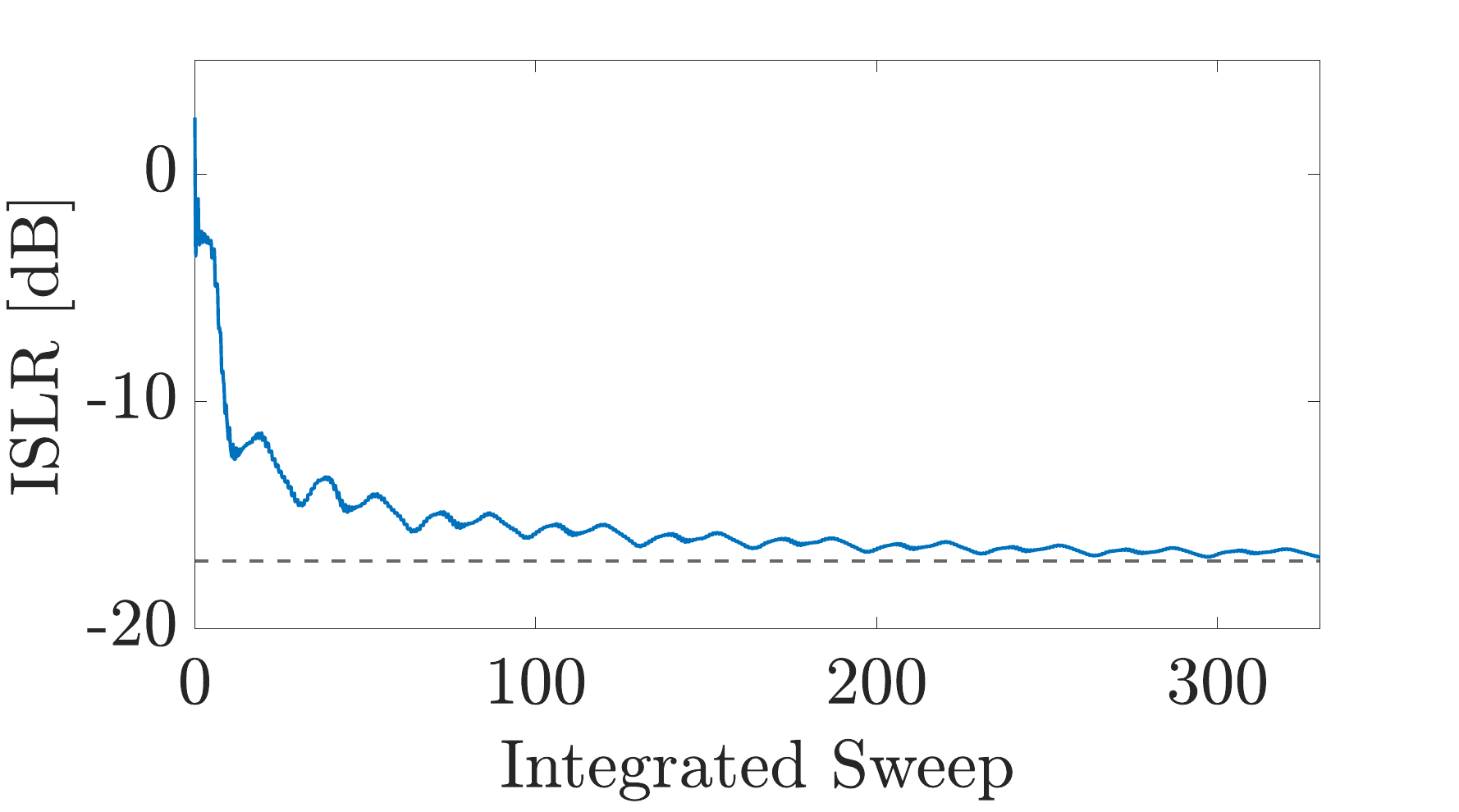}
    \caption{ISLR of a point target w.r.t. the number of integrated sweeps (blue), floor ISLR level for the considered ROI (dotted)}
    \label{fig:ISLR}
\end{figure}

Further, Fig. \ref{fig:ISLR} shows the integrated sidelobe ratio (ISLR) for a point target located in $\mathbf{r}^*$, as function of the number of employed Tx sweeps over the codebook $\Theta_i$. The ISLR is defined as
\begin{equation}
    \mathrm{ISLR} = \frac{ \iint \limits_{\mathcal{R} \setminus \Omega}|I(\mathbf{x})|^2 d\mathbf{x}}{ \iint \limits_{\Omega}|I(\mathbf{x})|^2 d\mathbf{x}}
\end{equation}
where $\mathcal{R} = \Delta_x \times \Delta_y = 1 \times 1$ m$^2$ is the ROI, while $\Omega = \delta_x \times \delta_y = 0.3 \times 0.15$ m$^2$ is the spatial region pertaining to the mainlobe of the spatial ambiguity function, whose size is defined in Appendix \ref{sec:resolution}. With the considered simulation settings, $|\Theta_i| = 60$ (number of beams per sweep) thus each sweep lasts for $T_\text{obs} = |\Theta_i| \Delta \tau = 3$ ms in time and correspond to a travelled distance of $v T_\text{obs} = 6$ cm. Fig. \ref{fig:ISLR} shows the ISLR of the resulting image obtained by summing the single images from each sweep.
The ISLR tends to converge to the nominal value, obtained for a reflection plane configured as a lens to focus the Tx signal in $\mathbf{r}^*$. The high order oscillations are due to the sparse array effect that can be observed in the exemplified raw data in Fig. \ref{fig:multisnap}. 

\begin{figure}
    \centering
    \subfloat[][$r_y=20$ m]{\includegraphics[width=0.35\columnwidth]{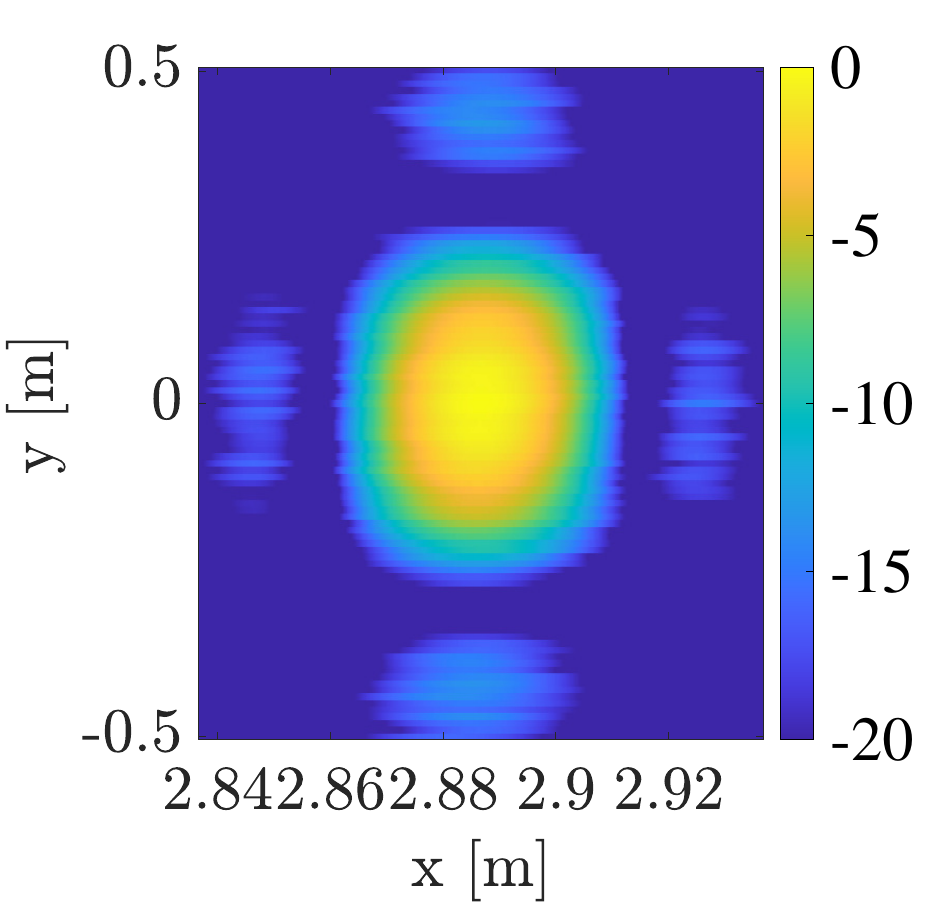}\label{subfig:20m}}
    \subfloat[][$r_y=15$ m]{\includegraphics[width=0.35\columnwidth]{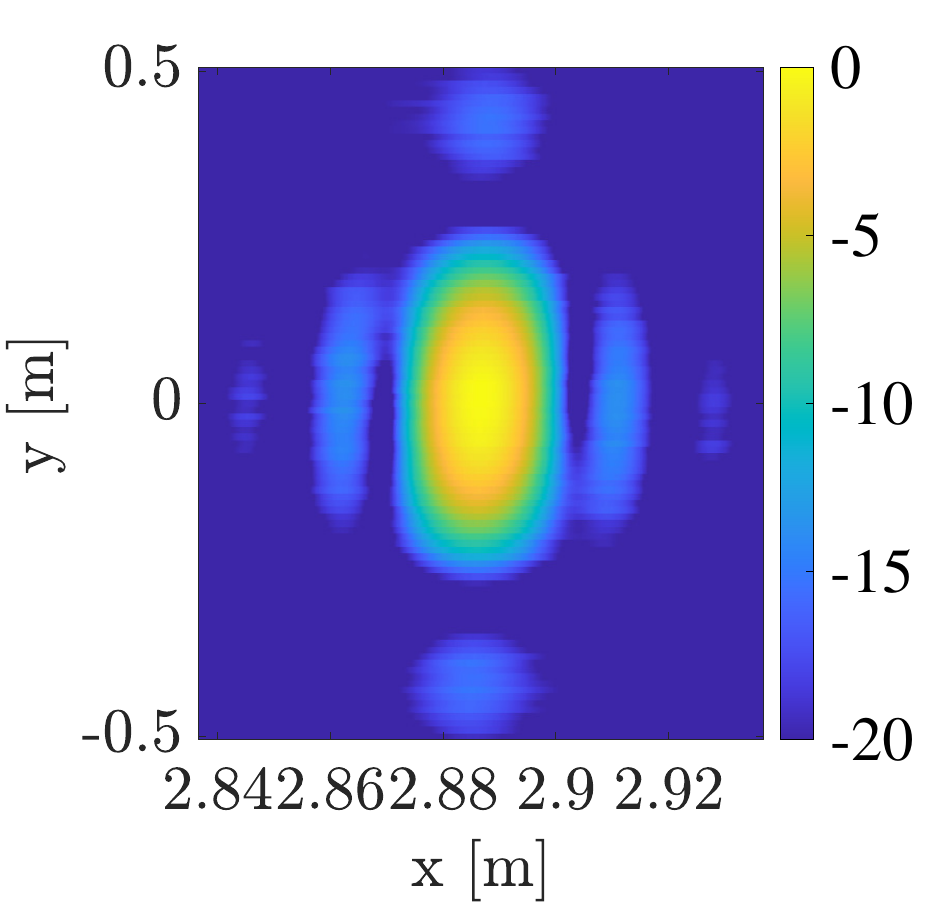}\label{subfig:15m}}\\ \vspace{-0.25cm}
    \subfloat[][$r_y=10$ m]{\includegraphics[width=0.35\columnwidth]{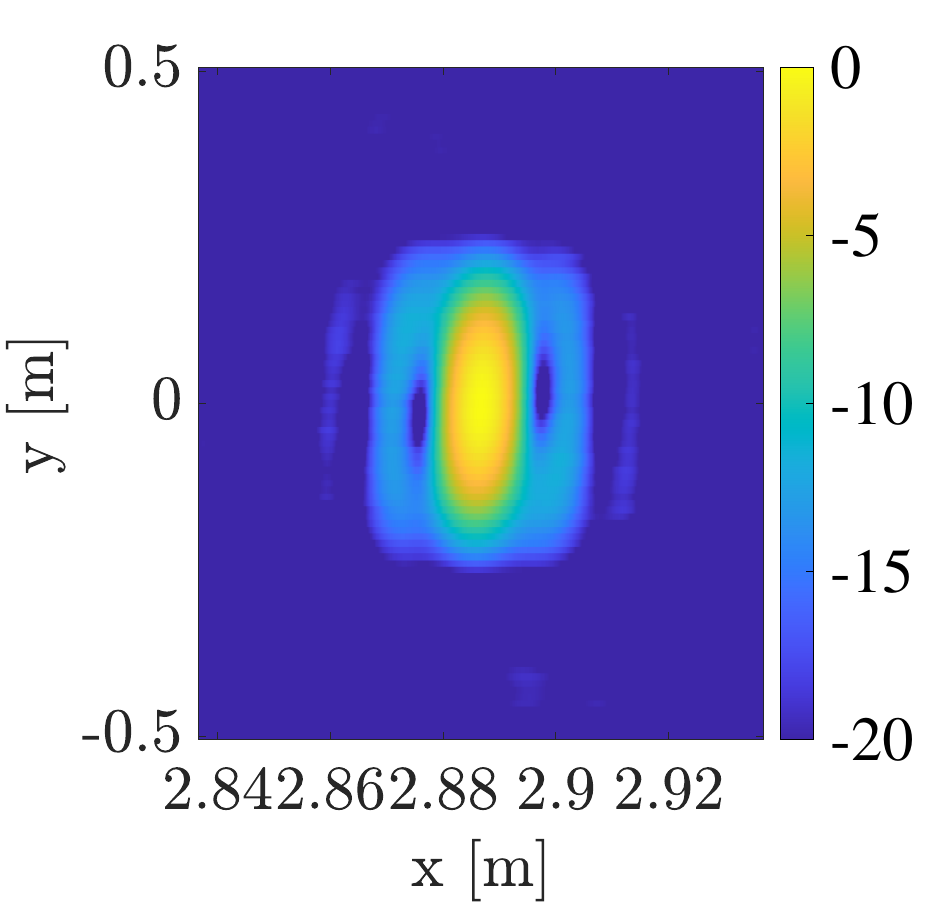}\label{subfig:10m}}
    \subfloat[][$r_y=5$ m]
    {\includegraphics[width=0.35\columnwidth]{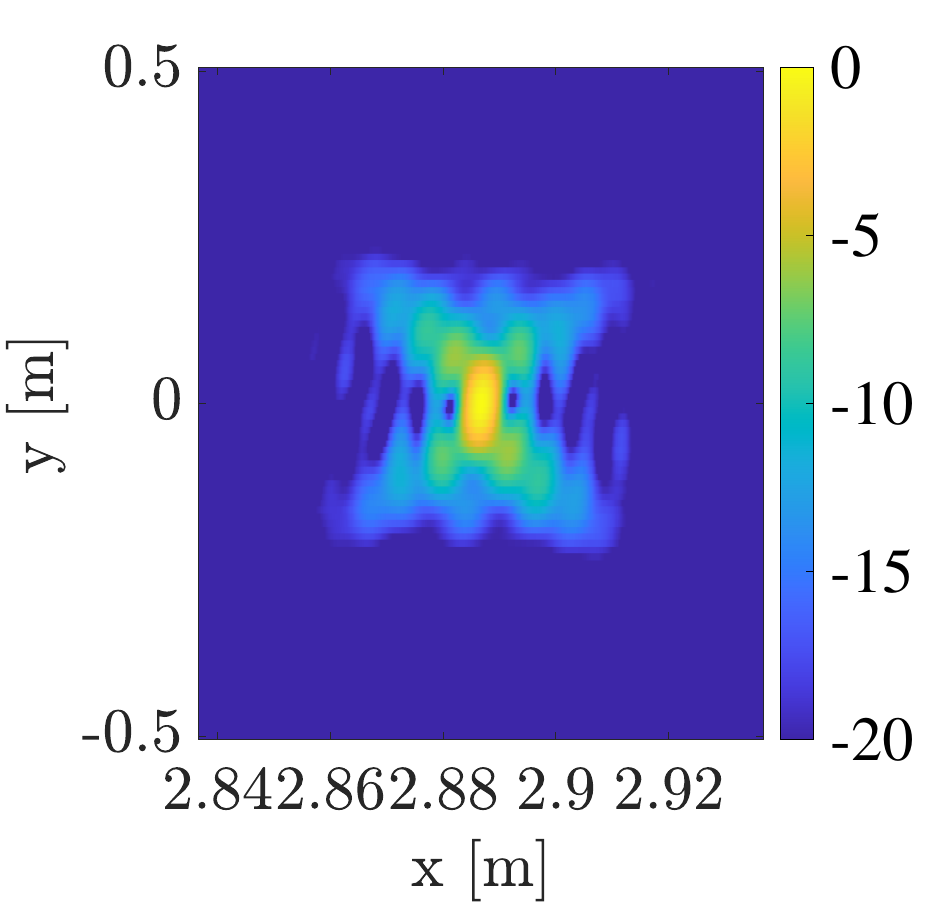}\label{subfig:5m}}
    \caption{Image of a point target varying the distance $r_y$ with the reflection plane. }
    \label{fig:range_res}
\end{figure}

As last result, in Fig. \ref{fig:range_res}, we show the image of a point target varying the distance with the reflection plane ($r_y$) from 20 m to 5 m. The result is obtained for $D=5$ m, $f_0=77$ GHz, $\overline{\theta}_i=30$ deg, $\Delta\theta_{i,\text{obs}}=15$ deg, $\Lambda=4$ m, $\overline{\theta}_o=0$ deg, such that to evaluate the enhancement of the range resolution, that in this case is along $y$. Reducing $r_y$, the proposed system allows improving not only the  cross-range resolution (along $x$, $\delta_x$) but also the range resolution $\delta_y$, thanks to the near-field effect. Notably, the proposed system benefits from the near-field effect without explicitly design the system accordingly; near-field sensing is achieved by successive far-field acquisitions. 

\section{Conclusion}\label{sec:conclusions}

This paper proposes a novel system for radio imaging in NLOS, with the aid of a properly configured anomalous reflection plane, that enables the illumination of the desired ROI in NLOS. The sensing terminal progressively illuminates contiguous portions of the plane---by sweeping a beam over a pre-defined codebook---while the plane implements a periodic angular reflection function to cover the ROI and increase the native resolution of the sensing terminal. We detail the system design criteria, discussing benefits, limitations and trade-offs. Moreover, we derive the effect of an imperfect knowledge of some selected system parameters, showing the effects on the generated image of a point target within the ROI. We demonstrate the superiority of our imaging method w.r.t. existing techniques (i.e., using a metallic mirror), that can attain the theoretical resolution limit in practical settings. Future activities involve the extension to the near-field operating condition and to the development of suitable countermeasures to the non-ideal knowledge of system parameters.


%

\appendices

\section{Theoretical vs. Effective Image Resolution}\label{sec:resolution}

The resolution of the stroboscopic imaging system is evaluated invoking diffraction tomography theory (DTT), stating that, under weak scattering from the targets (single bounce from each target), any radio image can be expressed either in Cartiesian coordinates or in the dual wavenumber domain~\cite{manzoni2023wavefield}. Let us recall the Rx signal model in Section \ref{sec:system_model}, where a source $\mathbf{s}$ illuminates a target in $\mathbf{r}$ via double reflection off the metasurface. Let us assume to transmit a single frequency $f = f_0 + f'$. Neglecting energy losses and noise (irrelevant herein), one can write the Rx signal as reported in \eqref{eq:Rxsignal_exttarget_RIS_doublebounce},
\begin{figure*}[!t]
\begin{equation}\label{eq:Rxsignal_exttarget_RIS_doublebounce}
\begin{split}
     y(\mathbf{s}, \mathbf{r},f)  & \approx \iint_{S}  \gamma(\mathbf{r}') \sum_{n\in \mathcal{M}} 
        \sum_{n'\in \mathcal{M}} e^{j\phi_n}e^{j\phi_{n'}} e^{-jk D_{\mathbf{s}n}} e^{-jk D_{n\mathbf{r}'}} e^{-jk D_{\mathbf{r}'n'}} e^{-jk D_{n'\mathbf{s}}} \mathrm{d}\mathbf{r'}\\
     & \overset{(a)}{\approx} \sum_{n\in \mathcal{M}} 
        \sum_{n'\in \mathcal{M}} e^{j\phi_n}e^{j\phi_{n'}} e^{-jk D_{\mathbf{s}n}} e^{-jk D_{n'\mathbf{s}}} e^{-jk D_{n\mathbf{r}}} e^{-jk D_{\mathbf{r}n'}} \iint_{S}  \gamma(\mathbf{r}' ) e^{-j k \left(\nabla D_{n \mathbf{r}} + \nabla D_{\mathbf{r}n'}\right)^T \mathbf{r'}}\mathrm{d}\mathbf{r'}
\end{split}
\end{equation}\hrulefill
\end{figure*}
where \textit{(i)} $S$ is the surface of the target, that is illuminated by the Tx signal, \textit{(ii)} $k = k_0 + k' = 2\pi f/c$ is the wavenumber for frequency $f$, \textit{(iii)} $\gamma(\mathbf{r}) = \sqrt{\sigma_\mathbf{r}}e^{j\psi}$ is the target's complex reflectivity, \textit{(iv)} $D_{\mathbf{s}n}=\|\mathbf{p}_n-\mathbf{s}\|$, $D_{n\mathbf{r}'}=\|\mathbf{r}'-\mathbf{p}_n\|$, $D_{\mathbf{r}'n'}=\|\mathbf{p}_{n'}-\mathbf{r}'\|$, $D_{n'\mathbf{s}}=\|\mathbf{s}-\mathbf{p}_{n'}\|$ are the involved distances (possibly approximated for far-field as in \eqref{eq:Rx_signal_generic_narrowbeam}). The Rx signal is the sum of all the contributions from each metasurface element, from each infinitesimal point on the target's surface. Approximation $(a)$ is for a Taylor expansion of distances around $\mathbf{r}$, where $\nabla D_{n \mathbf{r}}$ and $\nabla D_{\mathbf{r}n'}$ denote unit vectors pointing from the $n$th meta-atom to the target and from the target to the $n'$th meta-atom, respectively. 

Now, we recognize the integral in \eqref{eq:Rxsignal_exttarget_RIS_doublebounce}$(a)$ as the Fourier transform of the target's reflectivity evaluated in $\mathbf{k}_{nn'} = \mathbf{k}_{n \mathbf{r}} - \mathbf{k}_{\mathbf{r}n'}$, where
\begin{align}
    \mathbf{k}_{n \mathbf{r}} = \frac{2 \pi (f_0\hspace{-0.05cm}+\hspace{-0.05cm}f')}{c} \nabla D_{n \mathbf{r}}, \,\,
    \mathbf{k}_{\mathbf{r}n'} = -\frac{2 \pi (f_0\hspace{-0.05cm}+\hspace{-0.05cm} f')}{c} \nabla D_{\mathbf{r}n'}
\end{align}
denote plane wave vectors from the metasurface to the target and vice-versa, respectively.
Therefore, with the scattering from a single pair of metasurface elements ($nn'$), a single wavenumber $\mathbf{k}_{nn'} = \mathbf{k}_{n \mathbf{r}} - \mathbf{k}_{\mathbf{r}n'}$ is illuminated by the sensing imaging experiment. The set of illuminated wavenumbers, often called \textit{wavenumber coverage}, is obtained as a superposition of effects:
\begin{equation}\label{eq:wavenumber_cov_single}
    \mathcal{K}(\mathbf{s},\mathbf{r}|\mathcal{M},B) =  \left\{\mathbf{k}_{nn'} \bigg \lvert  n,n'\in \mathcal{M}, f' \in \left[-\frac{B}{2},+\frac{B}{2}\right]\right\}.
\end{equation}
The resolution in Cartesian coordinates (along $x$ and $y$) is defined as the width of the main lobe of the image of a point target located in $\mathbf{r}$. The resolution lower bound (thus the best performance) can be derived from the wavenumber coverage as follows:
\begin{equation}\label{eq:2Dresolution}
     \delta_x\approx \frac{2\pi}{\Delta k_x},\,\,\,\delta_y\approx \frac{2\pi}{\Delta k_y},
\end{equation}
where $\Delta k_x$ and $\Delta k_y$ denote the width of the wavenumber coverage along $x$ and $y$. Resolution limit \eqref{eq:2Dresolution} is the lower bound on the effective (achievable) resolution in the proposed system, that is only attained when the elements are configured to \textit{focus} the impinging signal from the source onto the target, i.e., 
\begin{equation}\label{eq:optim_config}
    \phi_n = 2 k_0 (D_{\mathbf{s}n} + D_{n \mathbf{r}}), \,\, \forall n\in \mathcal{M}.
\end{equation}
In this latter configuration, the SNR at the target's location is maximized and, most important, the resolution of the image is dictated by the portion of illuminated metasurface, inversely proportional to the source directivity. Indeed, decreasing the directivity of the source increases the number of illuminated elements $M$, thus the overall resolution. In this case the metasurface is used as a lens, and it works as an equivalent antenna array. In general, any configuration different from \eqref{eq:optim_config} leads to \textit{defocusing} of the target, thus loss of resolution and SNR, but the effective image resolution is ruled by the metasurface and by its phase configuration. The only exception is for a the metasurface configured to behave as a mirror (or anomalous mirror), such that $\phi_n = \alpha \, n \neq 2 k_0 (D_{\mathbf{s}n} + D_{n \mathbf{r}})$. In this latter case the resolution of the image is dictated by the source and \textit{not} by the metasurface.

When multiple ($P>1$) sets of meta-atoms $\mathcal{M}_p$, $p=1,...,P$ are illuminated and configured such that to reflect the impinging signal on the target $\mathbf{r}$, the wavenumber coverage is simply the union of sets as follows:
\begin{equation}\label{eq:wavenumber_cov_multiple}
     \mathcal{K}(\mathbf{s},\mathbf{r}|B) = \bigcup_{p} \,\mathcal{K}(\mathbf{s},\mathbf{r}|\mathcal{M}_p,B),
\end{equation}
and the resolution is evaluated as in \eqref{eq:2Dresolution}, plugging the compound set $\mathcal{K}(\mathbf{s},\mathbf{r}|B)$. Typically, if the $P>1$ sets $\mathcal{M}_p$ are characterized by different observation angles w.r.t. the target, the image resolution as defined in \eqref{eq:2Dresolution} increases. However, if the sets $\mathcal{M}_p$ are disjoint, the overall image is affected by significant sidelobes, due to the non-contiguous illumination of the metasurface.


\bibliographystyle{IEEEtran}
\bibliography{bibtex/bib/Bibliography}

\end{document}